\documentclass[conference]{IEEEtran}

\usepackage[normalem]{ulem}
\usepackage{subcaption}
\usepackage{graphicx}
\usepackage{listings}
\usepackage{xspace}
\usepackage{hyperref}
\usepackage{paralist}

\usepackage{algorithm}
\usepackage[noend]{algpseudocode}

\usepackage{amsmath}
\usepackage{amsthm}
\usepackage{amsfonts}
\usepackage[capitalize,nameinlink]{cleveref}
\usepackage{array}
\usepackage{booktabs}
\usepackage{boxedminipage}
\usepackage{calrsfs}
\usepackage{caption}
\usepackage{color}
\usepackage{enumitem}
\usepackage{fancyhdr}
\usepackage{float}
\usepackage{xcolor}
\usepackage{hyperref}
\hypersetup{
	unicode=true,
        bookmarksnumbered=true,
	colorlinks=true,
	linkcolor={red!70!black},
	citecolor={red!70!black},
	urlcolor={blue!70!black},
	pdfborder={0 0 0}
}
\usepackage[capitalize,nameinlink]{cleveref}
\usepackage[T1]{fontenc}
\usepackage[utf8]{inputenc}
\usepackage{microtype}
\microtypecontext{spacing=nonfrench}
\usepackage{multirow}
\usepackage{nicefrac}
\usepackage{pifont}
\usepackage{sidecap}
\usepackage{verbatim}
\usepackage{framed}
\usepackage{tcolorbox}
\usepackage{etoolbox}
\usepackage{MnSymbol}
\usepackage{flushend}
\usepackage{tikz}
\newcommand*\circled[1]{\tikz[baseline=(char.base)]{
            \node[shape=circle,draw,inner sep=2pt] (char) {#1};}}

\usepackage{cite}
\usepackage{textcomp}

\newif\ifshowcomment
\showcommenttrue

\ifshowcomment
\newcommand{\todo}[1]{\textcolor{blue}{[{TODO: #1}]}}

\newcommand{\pratyush}[1]{\textcolor{orange}{P: #1}}
\newcommand{\inigo}[1]{\textcolor{blue}{IG: #1}}
\newcommand{\esha}[1]{\textcolor{cyan}{EC: #1}}
\newcommand{\chaojie}[1]{\textcolor{teal}{CZ: #1}}
\newcommand{\aashaka}[1]{\textcolor{orange}{AS: #1}}
\else
\newcommand{\todo}[1]{}
\newcommand{\pratyush}[1]{}
\newcommand{\inigo}[1]{}
\newcommand{\esha}[1]{}
\newcommand{\chaojie}[1]{}
\newcommand{\aashaka}[1]{}
\fi

\newif\ifcolorrevisions
\colorrevisionsfalse

\ifcolorrevisions
    
    \newcommand{\rev}[1] {{\color{red}{#1}}}
\else
    
    \newcommand{\rev}[1]{#1}   
\fi

\newcommand{\papername}{Splitwise\xspace}

\usepackage{siunitx}

\newcommand{\myparagraph}[1]{\vspace{\smallskipamount}\noindent\textbf{#1.\xspace}}

\newcommand{\eg}{\emph{e.g.}\xspace}

\newcommand{\ie}{\emph{i.e.}\xspace}

\newcommand*{\rom}[1]{\uppercase\expandafter{\romannumeral #1\relax}}
\newcommand{\insightparagraph}[1]{\vspace{\smallskipamount}\noindent\textbf{\emph{Insight \rom{#1}:\xspace}}}

\def\BibTeX{{\rm B\kern-.05em{\sc i\kern-.025em b}\kern-.08em
    T\kern-.1667em\lower.7ex\hbox{E}\kern-.125emX}}

\title{\papername: Efficient Generative LLM Inference Using Phase Splitting}

\author{\\
Pratyush Patel$^1$,
Esha Choukse$^2$,
Chaojie Zhang$^2$,
\\
Aashaka Shah$^2$,
\'{I}\~{n}igo Goiri$^2$,
Saeed Maleki$^2$,
Ricardo Bianchini$^2$
\vspace{.15in}
\\
$^1$University of Washington\hspace{.4in}$^2$Microsoft 
}

\begin{document}

\maketitle
\thispagestyle{plain}
\pagestyle{plain}
\setcounter{page}{1}


\begin{abstract}

Generative large language model (LLM) applications are growing rapidly, leading to large-scale deployments of expensive and power-hungry GPUs.
Our characterization of LLM inference shows that each inference request undergoes two phases: a compute-intensive prompt computation phase and a memory-intensive token generation phase, each with distinct latency, throughput, memory, and power characteristics.
Despite state-of-the-art batching and scheduling, the token generation phase underutilizes compute resources.
Unlike prompt computation, token generation does not need the compute capability of the latest GPUs and can be run with lower power and cost. 

Based on these insights, we propose Splitwise, a model deployment and scheduling technique that splits the two phases of LLM inference requests on to separate machines.
Splitwise enables phase-specific resource management using hardware that is well suited for each phase.
Request state is transferred efficiently between machines using optimized network libraries on the fast back-plane interconnects available in today's GPU clusters.
Using Splitwise, we design homogeneous and heterogeneous LLM inference clusters optimized for throughput, cost, and power.
Compared to current designs, Splitwise clusters achieve up to $1.4\times$ higher throughput at 20\% lower cost. Alternatively, they can deliver $2.35\times$ more throughput under the same power and cost budgets.

\end{abstract}

\section{Introduction}
\label{sec:intro}

Recent advancements in generative large language models (LLMs) have significantly improved their response quality and accuracy~\cite{chatgptapi, hf_llama2}.
These trends have led to the widespread adoption of LLMs across various domains~\cite{bardassistant,newbing}.
Most modern LLMs are built using the transformer architecture~\cite{wolf2020transformers, vaswani2017attention} and exhibit similar characteristics~\cite{patel2024llmpower}.
Transformer model sizes have grown steadily, from the early BERT models~\cite{devlin2018bert} having 340 million parameters, to GPT-3~\cite{gpt3oai} with a staggering 175 billion parameters, and GPT-4 rumored to have even more.

LLMs typically run on expensive and power-hungry GPUs~\cite{dgxh100}.
The sudden and large-scale deployment of LLMs has led to a worldwide GPU capacity crunch~\cite{capacitycrunch}.
The computational demand for LLM inference far exceeds that of training due to the vast number of applications leveraging LLMs.
Furthermore, since training LLMs requires expensive and dedicated supercomputers~\cite{openai_7500k8s,meta_ai_supercluster}, a large number of inferences are necessary to amortize 
the high training costs.
LLM inference jobs, although orders of magnitude smaller than training, are still expensive given the compute involved.
\footnotetext[1]{Work partly done as an intern at Microsoft.}

\begin{table}[]
\centering
\footnotesize
\begin{tabular}{@{}cccc@{}}
\toprule
                       & \textbf{A100} & \textbf{H100}  & \textbf{Ratio}\\ \midrule
\textbf{TFLOPs}        & 19.5              & 66.9     & 3.43$\times$
\\
\textbf{HBM capacity}  & 80GB              & 80GB     & 1.00$\times$ \\
\textbf{HBM bandwidth} & 2039GBps          & 3352GBps & 1.64$\times$ \\
\textbf{Power}         & 400W              & 700W     & 1.75$\times$ \\
\textbf{NVLink}        & 50Gbps            & 100Gbps  & 2.00$\times$ \\
\textbf{Infiniband}    & 200GBps           & 400GBps  & 2.00$\times$ \\
\textbf{Cost per machine~\cite{coreweave}} & \$17.6/hr        & \$38/hr  & 2.16$\times$ \\ 
\bottomrule
\end{tabular}
\caption{NVIDIA A100 vs. H100 specifications.
}
\label{tab:a100_vs_h100_specs}
\vspace{-0.3cm}
\end{table}

Generative LLM inference for a single request consists of several forward passes through the model, since the output tokens are generated one by one.
This inherently has two contrasting phases of computation.
First, the \emph{prompt computation phase}, in which all the input prompt tokens run through the forward pass of the model in parallel to generate the first output token.
This phase tends to be computationally intensive and requires the high FLOPs (floating point operations per second) of the latest GPUs today.
Second, the \emph{token generation phase}, in which subsequent output tokens are generated sequentially based on the forward pass of the last token and all the cached context from previous tokens in the sequence.
Given the lack of compute parallelism, this phase tends to be more memory bandwidth and capacity bound, despite state-of-the-art batching.
Running both phases on the same machine \rev{often leads to inconsistent end-to-end latencies due to the arbitrary batching of prompt and token phases}.
Due to these challenges, services need to over-provision expensive GPUs to meet tight inference service level objectives (SLOs) for interactive applications.
At the same time, cloud service providers (CSPs) are having to build a lot of new datacenters to meet the GPU demand, and are running into a power wall~\cite{powerwall}. 

The industry continues to release new computationally powerful GPUs, each much more power hungry and expensive than the last.
However, as shown in \Cref{tab:a100_vs_h100_specs}, the high-bandwidth memory (HBM) capacity and bandwidth on these GPUs has not scaled at the same rate recently.
The latest NVIDIA H100 GPUs have $3.43\times$ more compute and $1.75\times$ more power compared to their predecessor A100 GPUs.
However, their memory bandwidth only grew by $1.6\times$, with no increase in memory capacity.

\myparagraph{Our work}
Given the distinct properties of prompt computation and token generation phases, we propose splitting the inference request and running them on separate machines.
Doing so allows us to separately manage hardware resources for each phase, thereby increasing the GPU utilization and the overall efficiency of the system.
It also enables using different, better-suited hardware for each phase.
To realize such a setup, the cached context from the prompt computation needs to be communicated over from the prompt processing machine to the token generation machine at low latency.
We implement these transfers in an optimized manner over the back-end Infiniband interconnects avaialble in datacenters today, allowing us to increase efficiency without any perceived performance loss. 

With Splitwise, we design clusters optimized for cost, throughput, and power, using production traces of LLM inference requests~\cite{azure_llm_trace}. 
Given the diverging memory and compute scaling rates across GPU generations, we also evaluate different GPUs and power caps for the different inference phases. 
This allows us to target better performance per dollar (Perf/\$) for users, and better performance per watt (Perf/W) for CSPs.
Additionally, users can target older GPUs, which are likely more readily available to them.

We show that Splitwise-based LLM inference clusters can achieve 1.4$\times$ higher throughput at 20\% lower cost than existing clusters. Alternatively, they can deliver 2.35$\times$ more throughput with the same cost and power budgets.

\myparagraph{Summary}
We make the following contributions:
\begin{enumerate}[leftmargin=*]
\item An extensive characterization of the differences in the execution and utilization patterns of the prompt and token generation phases in LLM inference on the NVIDIA A100 and H100 GPUs using production traces.
\item Splitwise, our technique for optimized utilization of available hardware, which splits the prompt computation and token generation phases onto separate machines.
\item A design exploration of homogeneous and heterogeneous cluster deployments with Splitwise to optimize the overall cost, request throughput, and provisioned power.
\item An evaluation of the systems designed with Splitwise using production traces.
\end{enumerate}

\section{Background}
\label{sec:background}

\subsection{Large Language Models}
Modern LLMs are based on transformers.
Transformer models use attention~\cite{vaswani2017attention} and multi-layer-perceptron layers to understand the inputs and generate an output, respectively.
Transformer-based LLMs include encoder-only~\cite{devlin2018bert,liu2019roberta}, decoder-only~\cite{radford2019gpt, scao2022bloom, hf_llama2}, and encoder-decoder~\cite{flan-t5_finetune} models. 
Generative LLMs, the focus of this paper, are usually either decoder-only, or encoder-decoder models.

\subsection{Generative LLM inference phases}
\label{sec:kvcachebg}
\begin{figure}[t]
    \centering
    \includegraphics[width=1\columnwidth]{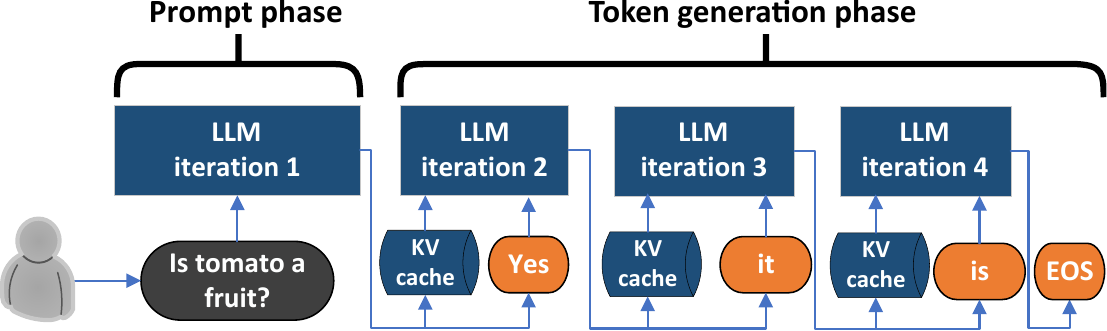}
    \caption{An LLM inference example.}
    \label{fig:LLM_example}
\end{figure}

\Cref{fig:LLM_example} shows an example of generative LLM inference.
Once the prompt query is received, all the input tokens are computed in parallel, within a single iteration, to generate the first token.
We call this the prompt processing phase.
The context generated from the attention layers during the prompt computation is saved in the key-value (KV) cache, since it is needed for all the future token generation iterations. 
After the first token is generated, the following tokens only use the last generated token and the KV-cache as inputs to the forward pass of the model.
This makes the subsequent token generation more memory bandwidth and capacity intensive than the computationally heavy prompt phase.


\subsection{Performance metrics for LLMs}
Prior work has proposed three main metrics for LLM inference: end-to-end (E2E) latency, time to first token (TTFT), and throughput.
We add another latency metric: time between tokens (TBT), to track the online streaming throughput of the tokens as they are generated serially.
\Cref{tab:perf_metrics} summarizes the key performance metrics that we consider in this work.

\begin{table}[t]
\centering
\footnotesize
\begin{tabular}{@{}ccc@{}}
\toprule
\textbf{Metric}            & \textbf{Importance to user} \\ \midrule
End-to-end (E2E) latency   & Total query time that the user sees        \\
Time to first token (TTFT) & How quickly user sees initial response    \\
Time between tokens (TBT)  & Average token streaming latency           \\
Throughput                 & Requests per second \\ \bottomrule
\end{tabular}
\caption{Performance metrics for LLMs.}
\label{tab:perf_metrics}
\end{table}

Generative LLMs may be used for a variety of tasks with different kinds of SLOs. 
For batch tasks (\eg, summarization), TTFT or TBT latency metrics are less important than throughput.
On the other hand, for latency-sensitive tasks (\eg, conversational APIs), TTFT and TBT are the more important metrics with tighter SLOs.

\begin{figure}[t]
    \centering
    \includegraphics[width=\columnwidth]{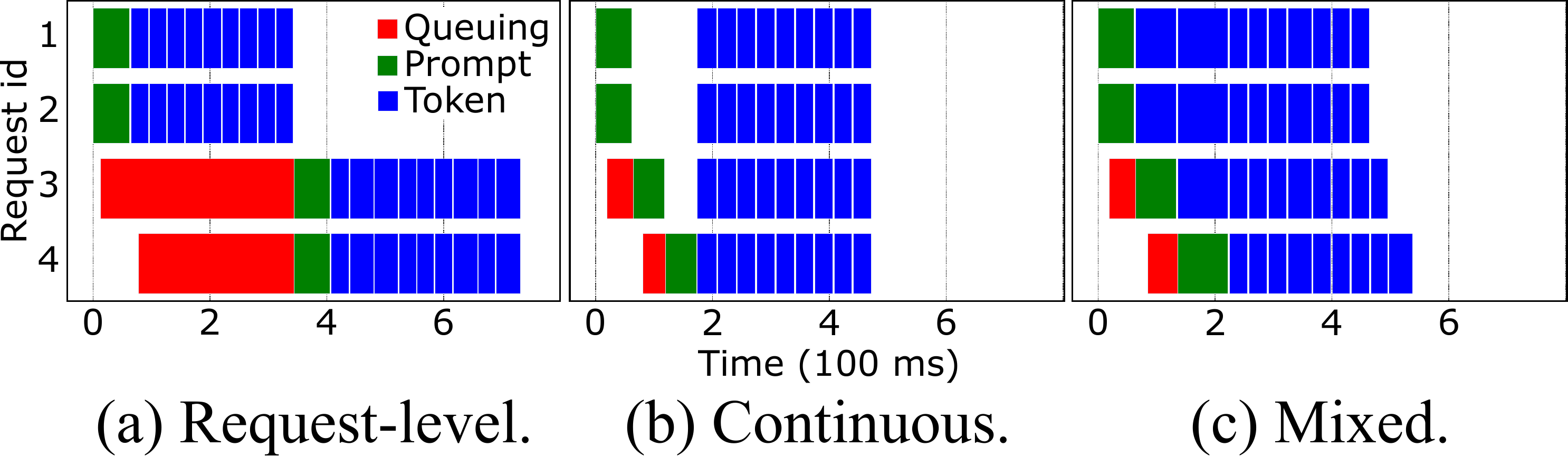}
    \caption{Batching mechanisms and their latency impact on the prompt and token phases.
    }
    \label{fig:batching-gantt-chart}
\end{figure}

\subsection{Batching of requests}
Inference requests can be batched together for higher throughput.
Several prior works have explored batching~\cite{yu2022orca,sarathi}.
\Cref{fig:batching-gantt-chart} shows the timelines for inference with three common batching mechanisms.
The default mechanism only batches at the \emph{request-level} (\Cref{fig:batching-gantt-chart}(a)).
In this case, ready requests are batched together, but all the forward passes for these requests are completed before any other requests are run.
Since requests can have long token generation phases, this can lead to long wait times for requests arriving in between, causing high TTFT and high E2E latencies.
An optimization is \emph{continuous batching}~\cite{yu2022orca} (\Cref{fig:batching-gantt-chart}(b)).
In this case, scheduling decisions are made before each forward pass of the model. However, any given batch comprises either only of requests in their prompt phase or only requests in token phase.
Prompt phase is considered more important since it impacts TTFT.
Hence, a waiting prompt can preempt a token phase.
Although this leads to shorter TTFT, it can substantially increase the tail for TBT, and therefore the E2E latency.
Finally, there is \emph{mixed batching} (\Cref{fig:batching-gantt-chart}(c))~\cite{sarathi}.
With this batching, the scheduling decisions are made at each forward pass, and the prompt and token phases can run together.
This reduces the impact on TBT, but does not eliminate it, since token phases scheduled with prompt phases will experience a longer runtime.
In the rest of the paper, we use mixed batching unless stated otherwise.

\subsection{Model parallelism}
\label{sec:modelparallel}
Model parallelism can be used to divide a model onto multiple GPUs, and even multiple machines, for higher efficiency and memory capacity.
LLM inference typically uses pipeline and tensor parallelism.
Pipeline parallelism (PP) divides the layers of the model among the GPUs, while keeping all the operators and tensors within a layer on the same GPU.
Tensor parallelism (TP) divides the tensor across the GPUs, while replicating all the layers on each GPU.
Pipeline parallelism requires lower communication across the participating GPUs, while tensor parallelism requires high bandwidth communication for each layer. 
In general, tensor parallelism performs better for GPUs within the same machine, connected with high bandwidth interconnects like \eg NVLink~\cite{nvidia_dgx_a100}.
In the rest of the paper, we use tensor parallelism across 8 GPUs for the best latency.

\subsection{GPU clusters and interconnects}
\label{sec:gpu-interconnect}
With the recent rise of LLM use cases, several cloud service providers have expanded the GPU-based offerings, leading to large GPU cluster deployments~\cite{meta_ai_supercluster,azure_openai,coreweave}.
Each machine in these AI clusters is generally comprised of 8 flagship NVIDIA GPUs (A100 or H100).
Each GPU is connected to all the other GPUs in the cluster with a high bandwidth Mellanox InfiniBand interconnect~\cite{infiniband,azure_ib}, forming a high bandwidth data plane network.
The InfiniBand bandwidth offered in the cloud today ranges from 25 to 50GBps per GPU pair~\cite{azure_ib,coreweave_ib}.

\section{Characterization}
\label{sec:motivation}
In this section, we explore the performance and utilization characteristics of LLM inference and draw key insights to guide the design of Splitwise. 

\myparagraph{Production traces}
We use production traces taken from two Azure LLM inference services on November 11$^{th}$ 2023. 
Our traces represent the most common scenarios in LLM inference today:
\emph{coding} and \emph{conversation}.
We have released a subset of our traces at \url{https://github.com/Azure/AzurePublicDataset}~\cite{azure_llm_trace}.
The traces we use for characterization are 20 minutes long and include the arrival time, input size (number of prompt tokens), and output size (number of output tokens).
Due to customer privacy requirements (\eg, GDPR), we do not have visibility into the content of the prompts.
We instead use the production traces to guide the input and output sizes, where we send the input prompt with the required number of tokens, 
and force the model to generate the corresponding number of output tokens for each request.
Note that the text of the inputs prompts does not impact the performance metrics that we benchmark, since they depend only on the input and output sizes.
For this characterization, we do not reuse the KV-cache between requests to emulate a cloud service with security guarantees.

\myparagraph{Models}
\Cref{tab:models} shows the models that we evaluate.
Both BLOOM~\cite{scao2022bloom} and Llama2~\cite{hf_llama2} are state-of-the-art open source LLMs.
Both models are decoder-only, transformer-based models. 
We use the version of each model with the most parameters, since these versions are the most representative for production-class accuracy.
Unless stated otherwise, we run BLOOM-176B and Llama-70B on vLLM~\cite{vllm} on a machine with 8 H100~\cite{dgxh100} GPUs.

\begin{table}[t]
\centering
\footnotesize
\begin{tabular}{@{}cccc@{}}
\toprule
\textbf{Model}      & \textbf{\#Layers} & \textbf{Hidden size} & \textbf{\#Heads} \\ \midrule
\textbf{Llama2-70B} & 80                &    8192              & 32                \\
\textbf{BLOOM-176B} & 70                & 14336                & 112              \\ \bottomrule
\end{tabular}
\caption{Models we evaluate and their parameters.}
\label{tab:models}
\end{table}

\subsection{Number of prompt and generated tokens}
\begin{figure}
    \centering
    \subfloat[Prompt input tokens.]{
        \includegraphics[width=0.48\columnwidth]{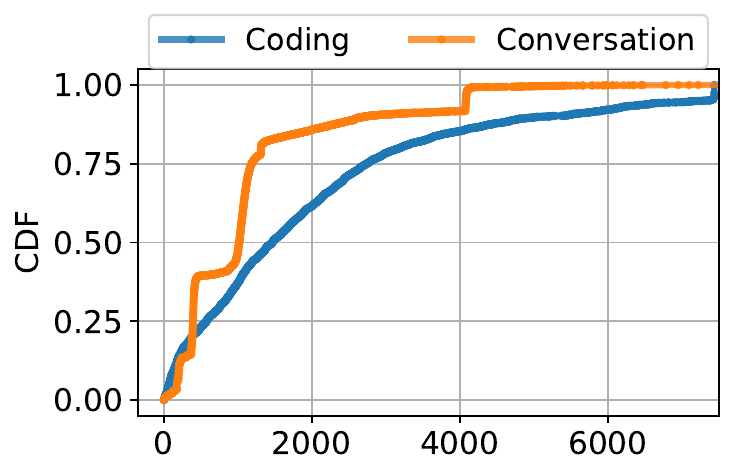}
        \label{fig:prompt_sizes}}
    \subfloat[Generated output tokens.]{
        \includegraphics[width=0.48\columnwidth]{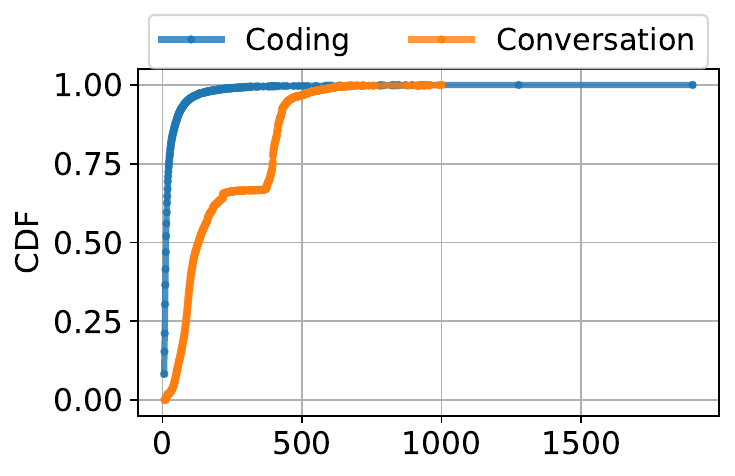}
        \label{fig:token_sizes}}
\caption{Distribution for prompt and generated tokens.}
    \label{fig:prompt_token_sizes} 
\end{figure}

To better understand our traces, we examine the distribution of the number of \emph{prompt input} and \emph{generated output} tokens.
\Cref{fig:prompt_sizes} shows the distribution of number of prompt tokens.
Since the coding LLM inference service is generally used to generate completions as the user is writing code, its input prompt can include large chunks of the code written so far.
Thus, it has a large median prompt size of 1500 tokens.
On the other hand, the conversation service has a wider range of input prompt tokens since it depends on the user.
The median number of prompt tokens for this trace is 1020 tokens.

\Cref{fig:token_sizes} shows the distribution of the number of generated tokens.
Since the coding service typically only generates the next few words in the program as the user types, the median number of output token is 13 tokens.
On the other hand, the conversation service has an almost bimodal distribution, with a median of 129 tokens generated.

\insightparagraph{1}
Different inference services may have widely different prompt and token distributions.

\subsection{Batch utilization}

\begin{figure}
    \centering
    \includegraphics[width=0.7\columnwidth]{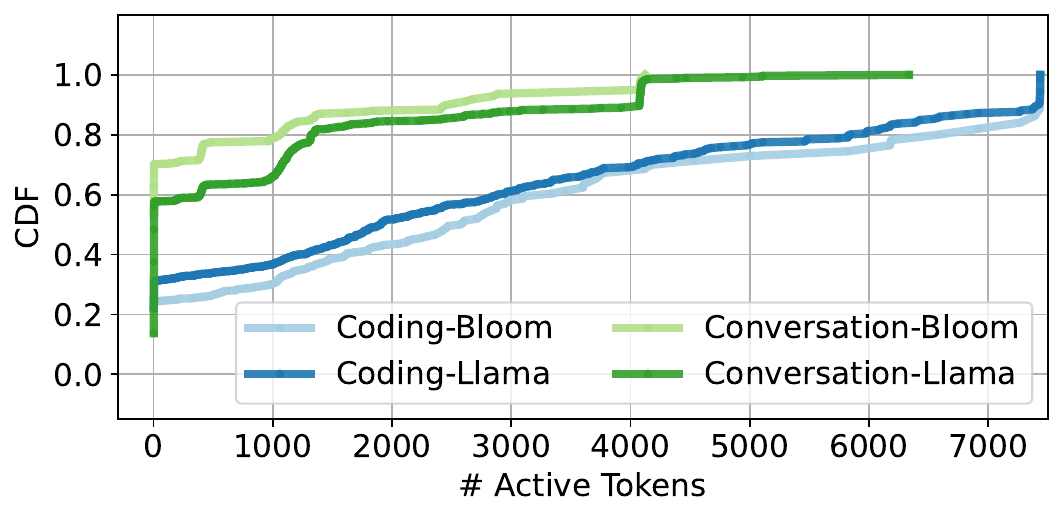}
    \caption{Cumulative distribution of time spent with various active batched tokens.}
    \label{fig:pvt-distribution}
\end{figure}

To understand how much can these requests be batched, we measure how often machines run at a given batch size.
We use mixed continuous batching as shown in \Cref{fig:batching-gantt-chart}.
To fit into a single machine, we run a scaled-down version of the coding and conversation traces with 2 requests per second.

\Cref{fig:pvt-distribution} shows the distribution of the time spent by the machine running various number of active tokens in a batch.
Note that if a prompt of 100 tokens is running in its prompt phase, we count the active tokens as 100.
However, once the request is in the token phase, we count it as one active token, since the tokens are generated one at a time (assuming a beam search size of one~\cite{vllm}).
We find that most of the time (60--70\%) for conversation is spent running only 20 tokens or fewer. 
Since the coding service has very few output tokens, it experiences even worse batching in the token phase and runs with a single token for more than 20\% of the time.
Both the LLMs show very similar trends.

\insightparagraph{2}
Mixed continuous batching spends most of the time with very few active tokens batched.

\begin{figure}
    \centering
    \subfloat[TTFT by prompt size.]{
        \includegraphics[width=0.32\columnwidth]{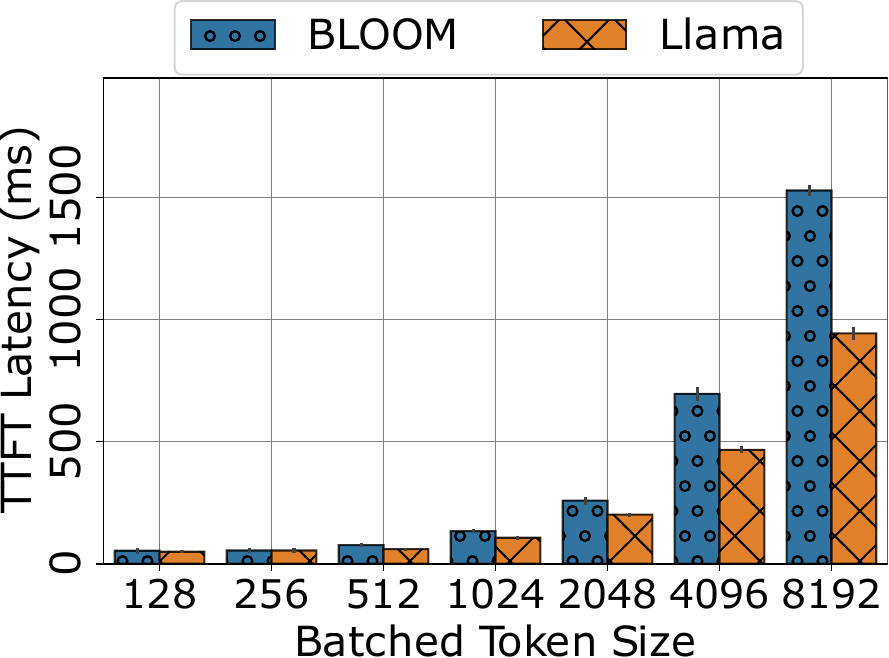}
        \label{fig:characterization_TTFT}}
    \subfloat[TBT by batch size.]{
        \includegraphics[width=0.32\columnwidth]{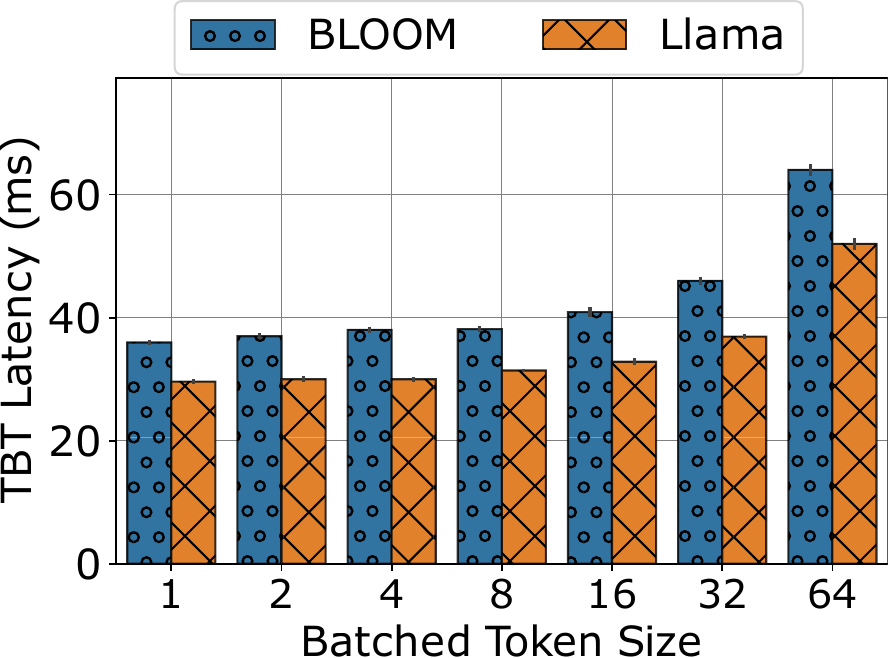}
        \label{fig:characterization_TBT}}
    \subfloat[Latencies on prod traces (no batching).]{
        \includegraphics[width=0.32\columnwidth]{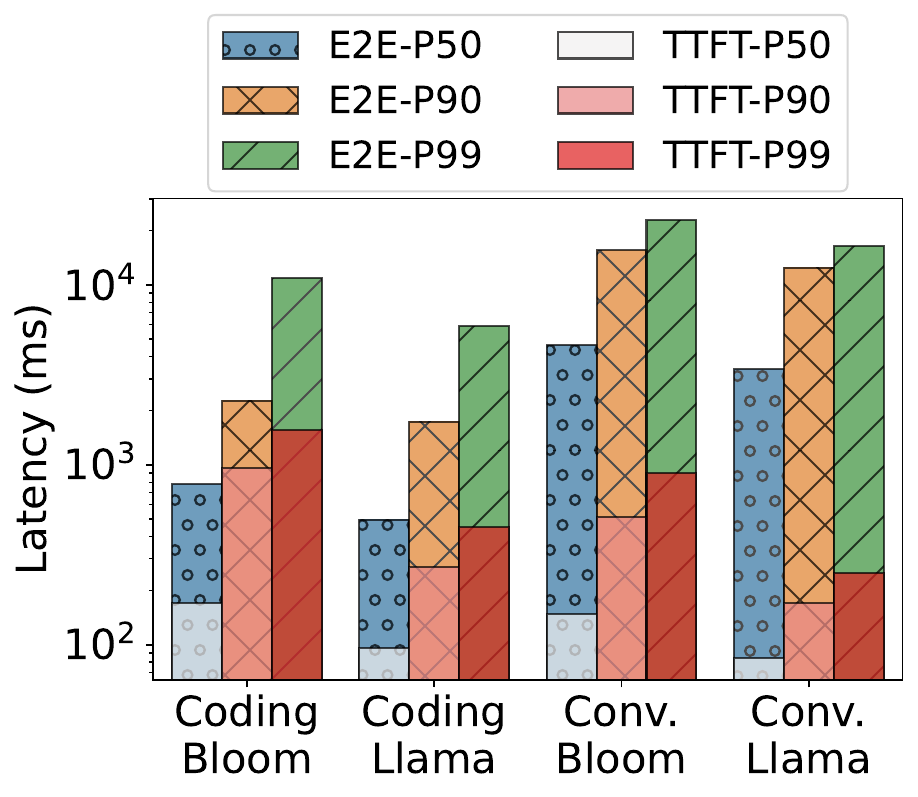}
        \label{fig:characterization_E2E}}            
\caption{TTFT, TBT, and E2E for BLOOM-176B and Llama-70B on DGX-H100.}
    \label{fig:char_TTFT_TBT_E2E} 
\end{figure}

\subsection{Latency}

\myparagraph{TTFT}
\Cref{fig:characterization_TTFT} shows the impact of the number of prompt tokens on TTFT.
The range of sizes was chosen based on the coding and conversation traces.
We find that TTFT for both models grows almost linearly as the prompt size increases.
This behavior is due to the prompt phase having high GPU utilization and being computationally bound.

\myparagraph{TBT}
\Cref{fig:characterization_TBT} shows the impact of forcefully batching the output tokens of different requests together on the TBT.
We observe very little impact on TBT as the batch size grows.
With a batch size of 64, there is only $2\times$ impact on TBT.

\myparagraph{E2E}
\Cref{fig:characterization_E2E} shows various percentiles of E2E latency for both models, with no batching.
The variability between the request input and output sizes is apparent.
Furthermore, we see that most of the E2E time is spent running the token phase.
This holds true even for the coding trace, where prompt sizes are large and generated tokens few. 
In fact, we find that for BLOOM-176B, a prompt phase with 1500 input tokens takes the same time as token phase with only 6 output tokens.

\insightparagraph{3}
For most requests, the majority of the E2E time is spent in the token generation phase.

\begin{figure}
    \centering
    \subfloat[Prompt phase.]{
        \includegraphics[width=0.48\columnwidth]{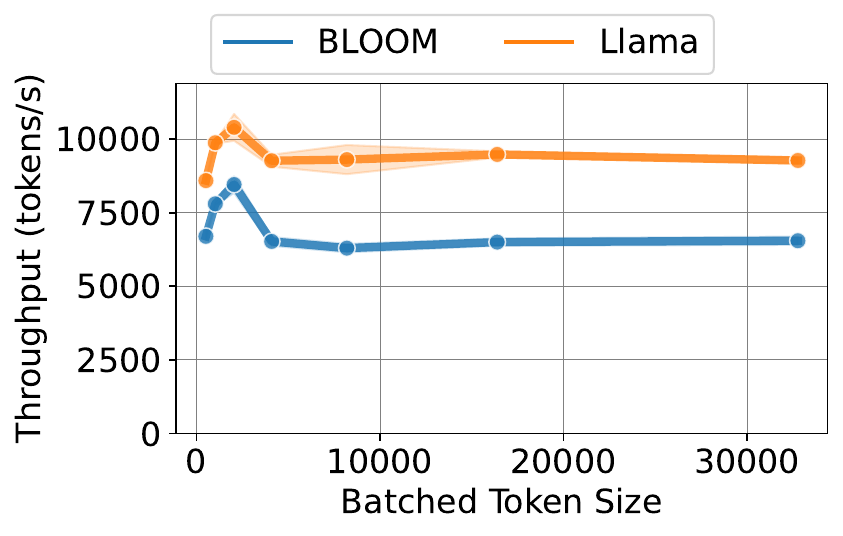}
        \label{fig:char_tput_prompt}}
    \subfloat[Token generation phase.]{
        \includegraphics[width=0.48\columnwidth]{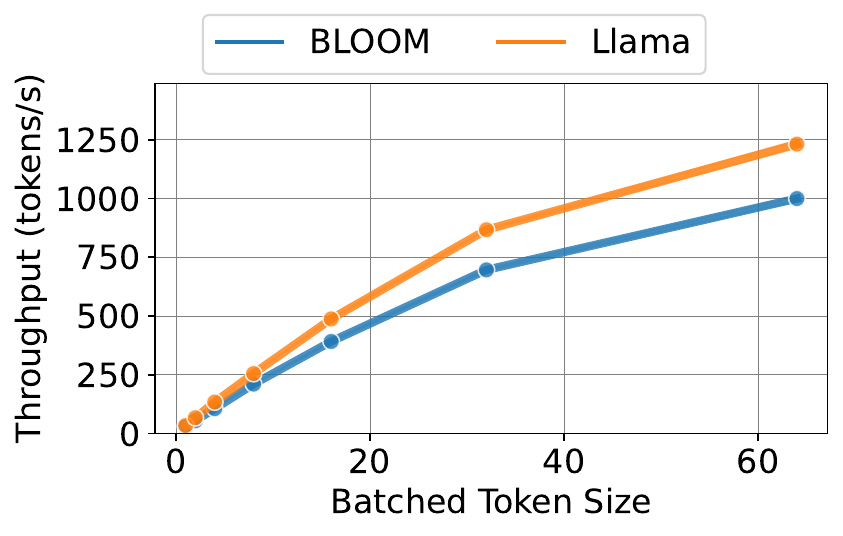}
        \label{fig:char_tput_token}}
    \caption{Impact of batching on the throughput for the 2 LLMs.}
    \label{fig:char_tput} 
\end{figure}

\subsection{Throughput}
\Cref{fig:char_tput} shows the impact of batching on the throughput (measured as tokens per second).
For the prompt phase, we define the throughput as the number of prompt input tokens that are processed per second.
We see that the throughput decreases after 2048 prompt tokens, which corresponds to a batch size of less than 2 for the median prompt sizes from the traces.
On the other hand, \Cref{fig:char_tput_token} shows that the throughput in the token phase keeps increasing with batching until 64 batch-size, at which point, the machine runs out of memory.

\insightparagraph{4}
The prompt phase batch size should be limited to ensure good performance.
In contrast, batching the token generation phase yields high throughput without any downside.

\begin{figure}
    \centering
     \includegraphics[width=0.60\columnwidth]{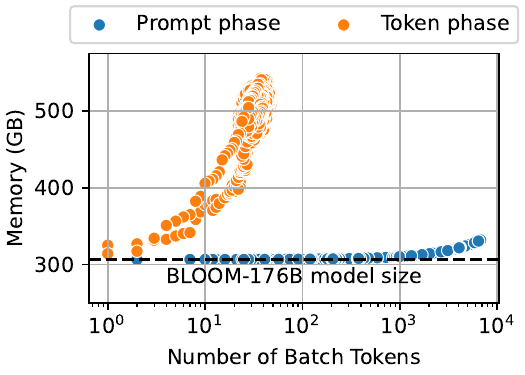}
    \caption{Required memory with batching in prompt/token phases.}
    \label{fig:char_memory}
\end{figure}

\subsection{Memory utilization}
During an LLM inference, the GPU memory is used to host the model weights and activations, as well as the KV caches (\Cref{sec:kvcachebg}).
As the number of tokens in a batch increase, the memory capacity required for the KV cache also increases.
\Cref{fig:char_memory} shows the memory capacity utilization during each phase as the number of tokens in the batch increases.
During the prompt phase, the input prompt tokens generate the KV cache.
During the output token phase, \emph{each} active generated token that is being processed accesses the KV cache of its \emph{entire context} so far.

\insightparagraph{5}
Batching during the prompt phase is compute-bound, whereas the token phase is limited by memory capacity.

\begin{figure}
    \centering
    \subfloat[Prompt phase.]{
        \includegraphics[width=0.48\columnwidth]{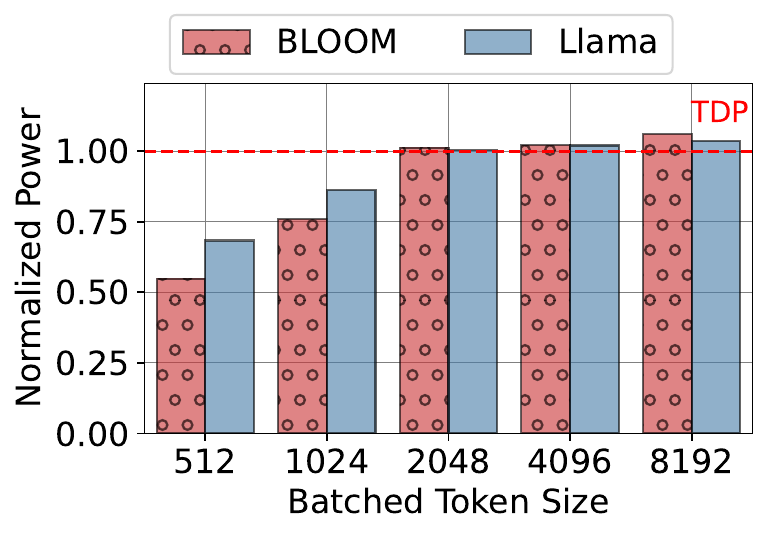}
        \label{fig:char_power_prompt}}
    \subfloat[Token generation phase.]{
        \includegraphics[width=0.48\columnwidth]{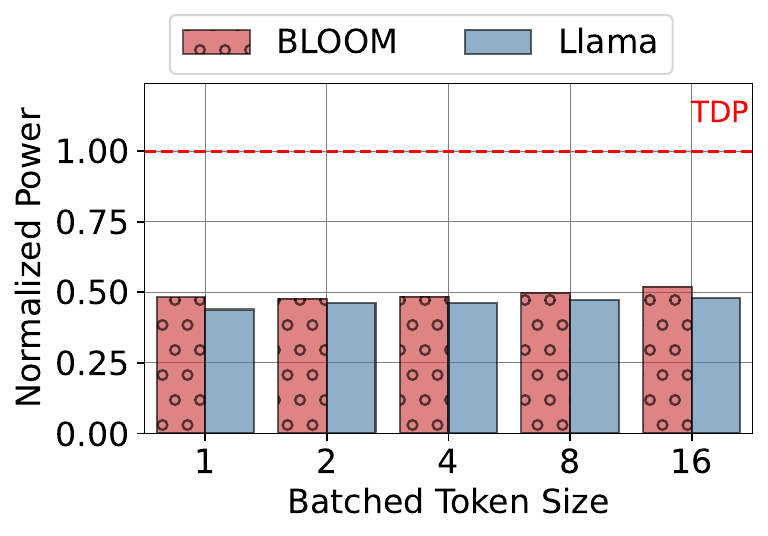}
        \label{fig:char_power_token}}
    \caption{Maximum and mean power utilization varying the batching size.
    }
    \label{fig:char_power}
\end{figure}

\begin{figure}[t]
    \centering
    \subfloat[Prompt phase.]{
        \includegraphics[width=0.48\columnwidth]{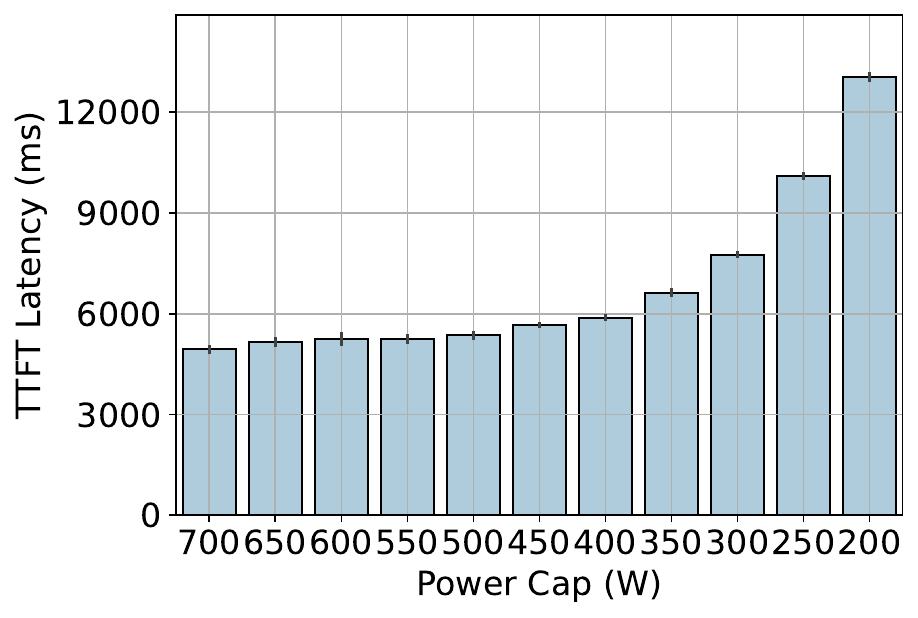}
        \label{fig:char_power_cap_prompt}}
    \subfloat[Token generation phase.]{
        \includegraphics[width=0.48\columnwidth]{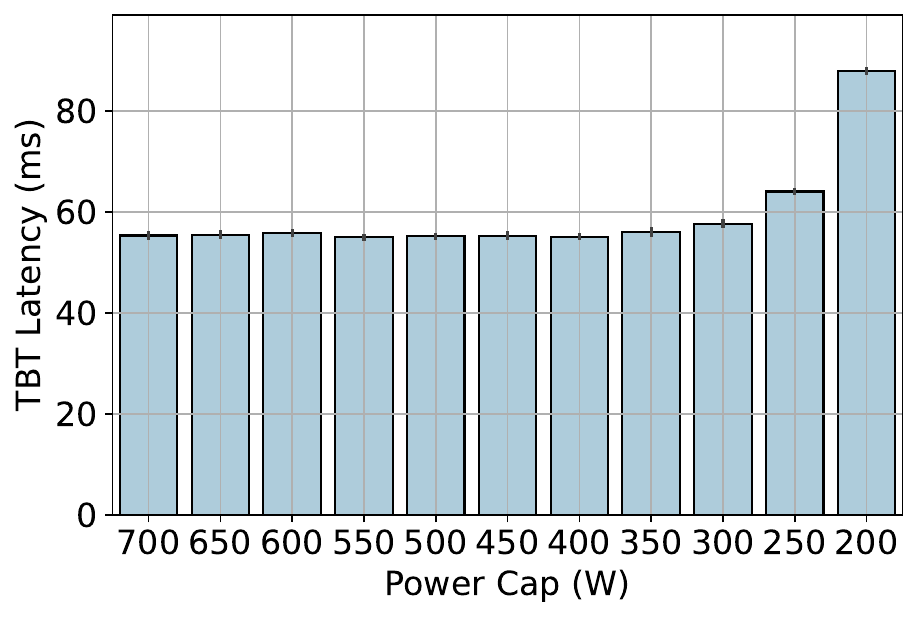}
        \label{fig:char_power_cap_token}}
    \caption{Impact of power cap on the prompt and token generation latency with the maximum batch size possible.
    }
    \label{fig:char_powercap}
\end{figure}

\subsection{Power utilization}
When hosting machines, cloud providers need to consider the peak power draw, which has direct impact in the datacenter cost~\cite{barroso2018_Datacenter}.
This is especially important when building GPU clusters, since GPUs consume much higher power than regular compute machines~\cite{patel2024llmpower,patel2023towards}.
\Cref{fig:char_power} shows the GPU power draw normalized to the thermal design power (TDP) when running prompt and token generation phases.
Since the the prompt phase is compute intensive, its power draw increases with batch size.
On the other hand, the token phase is memory bound and its power draw does not vary when increasing the number of tokens to process.

Providers can cap the power usage of the machines to reduce the peak power.
\Cref{fig:char_powercap} shows the impact to latency when increasing the power caps for both prompt and token phases.
The prompt phase is highly sensitive to the power cap and the latency increases substantially.
On the other hand, the token generation phase incurs almost no latency impact when power capping by over 50\% (\ie, 700 to 350W).

\insightparagraph{6}
While the prompt phase utilizes the power budget of the GPU efficiently, the token phase does not.

\begin{table}[t]
\centering
\footnotesize
\begin{tabular}{@{}ccccccc@{}}
\toprule
                      &\multicolumn{3}{c}{\textbf{Coding}} & \multicolumn{3}{c}{\textbf{Conversation}}\\
                      & \textbf{A100}   &\textbf{H100}  &\textbf{Ratio}      &\textbf{A100}      &\textbf{H100} &\textbf{Ratio}\\\midrule                      
\textbf{TTFT}         & 185 ms   & 95 ms  & 0.51$\times$ & 155 ms  & 84 ms   & 0.54$\times$ \\
\textbf{TBT}          & 52 ms    & 31 ms  & 0.70$\times$ & 40 ms   & 28 ms   & 0.70$\times$ \\
\textbf{E2E}          & 856 ms   & 493 ms & 0.58$\times$ & 4957 ms & 3387 ms & 0.68$\times$ \\
\textbf{Cost~\cite{coreweave}} & \$0.42 & \$0.52 & 1.24$\times$ & \$2.4  & \$3.6 & 1.5$\times$ \\
\textbf{Energy}       & 1.37 Whr & 1.37 Whr & 1$\times$  & 7.9 Whr   & 9.4 Whr    & 1.2$\times$ \\
\bottomrule
\end{tabular}
\caption{P50 request metrics on A100 vs. H100 without batching on Llama-70B.}
\label{tab:a100h100results}
\end{table}

\subsection{GPU hardware variations}
Given the different characteristics of prompt and token generation phases, we measure the performance impact on the two from running on different hardware.
\Cref{tab:a100_vs_h100_specs} shows the specifications for DGX-A100~\cite{nvidia_dgx_a100} and DGX-H100~\cite{dgxh100}. The memory-to-compute ratio favors A100 over H100.
Table~\ref{tab:a100h100results} shows our findings. 
We see a lower performance impact on the token generation phase (TBT) as compared to the Prompt phase (TTFT).
Since coding requests are dominated by prompt phase, by having very few generated tokens, the E2E latency impact from A100 is worse on coding than conversation.
Furthermore, we see that A100 has better or equal inference cost and energy overall compared to H100.

\insightparagraph{7}
Token generation can be run on less compute-capable hardware for better Perf/W and Perf/\$ efficiencies.

\section{\papername}

Based on our characterization insights, we propose Splitwise, a technique to split the prompt and generation phases in the LLM inference on to separate machines.

\begin{figure}
    \centering
    \includegraphics[width=0.8\columnwidth]{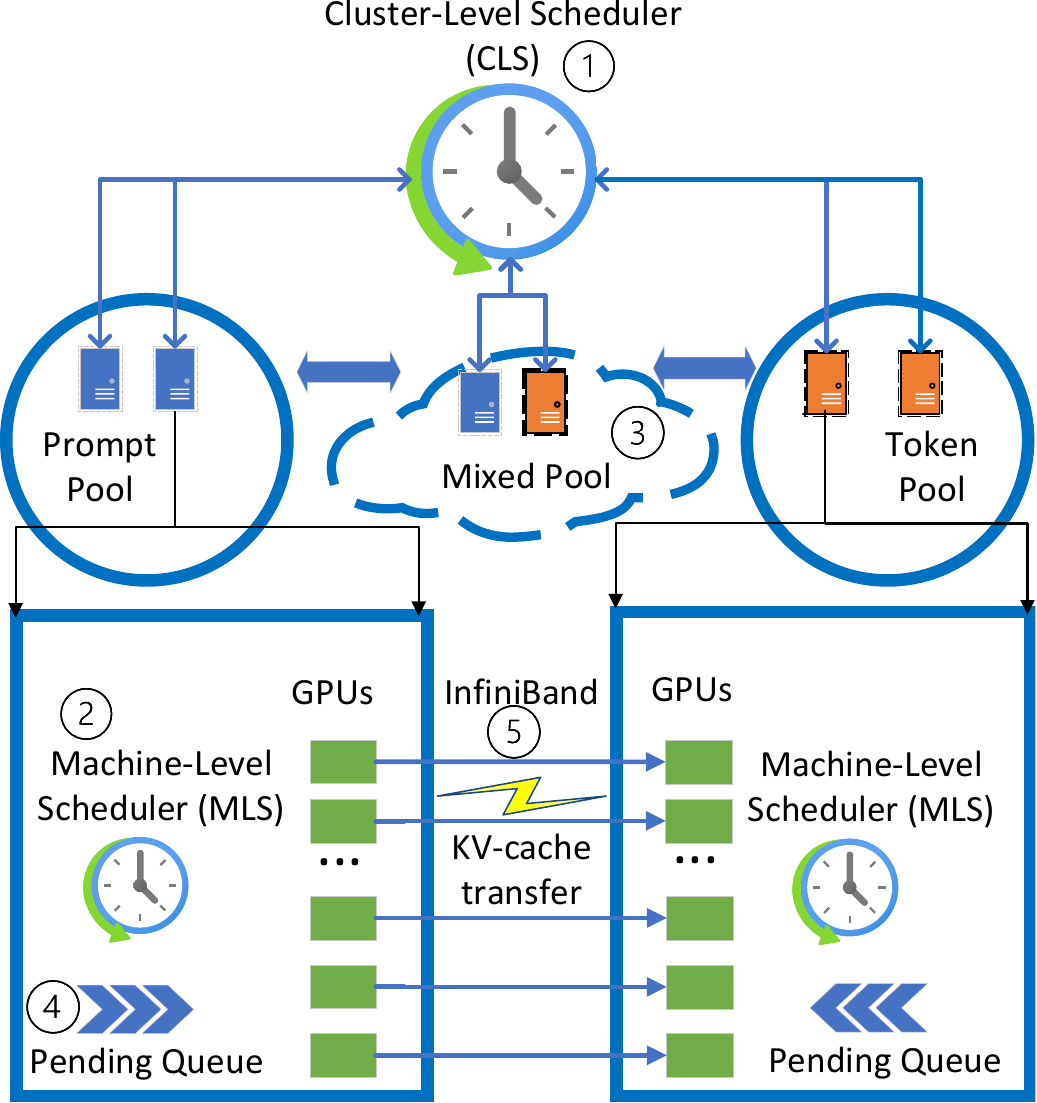}
    \caption{High-level system diagram of Splitwise.
    }
    \label{fig:systemdiag}
\end{figure}

\Cref{fig:systemdiag} shows the high-level overview of Splitwise. 
We maintain two separate pools of machines for prompt and token processing. 
A third machine pool, the mixed pool, expands and contracts as needed by the workload.
All machines are pre-loaded with the model of choice. 
When a new inference request arrives, the scheduler allocates it to a pair of machines (\ie, prompt and token).
The prompt machines are responsible for generating the first token for an input query, by processing all the input prompt tokens in the prompt phase and generating the KV-cache. 
The prompt machine also sends over the KV-cache to the token machine, which continues the token generation until the response is complete.
We use continuous batching at the token machines to maximize their utilization.
Machines in mixed pool use mixed continuous batching.

At a lower request rate, we target better latency in Splitwise, while, at a higher request rate, we target avoiding any performance or throughput reduction due to the fragmentation between prompt and token machine pools.

Splitwise uses a hierarchical two-level scheduling as shown in \Cref{fig:systemdiag}.
The cluster-level scheduler (CLS) \circled{1} is responsible for machine pool management and for routing incoming inference requests.
The machine-level scheduler (MLS) \circled{2} maintains the pending queue and manages batching of requests at each machine.

\subsection{Cluster-level scheduling}
\label{sec:scheduler}

\myparagraph{Machine pool management}
The CLS maintains the prompt, token, and mixed machine pools \circled {3}.
Splitwise initially assigns machines to the prompt or token pool depending on the expected request load and input/output token distributions.
Machines from the prompt or token pools may be dynamically moved into and out of the mixed pool to reduce fragmentation and meet SLOs at higher loads. 
A machine in the mixed pool retains its identity as a prompt or token machine and goes back to its original pool once there are no tasks of the opposite kind in its pending queue.
Switching pools does not incur any noticeable latency.
If the load distribution deviates considerably from initial assumptions, Splitwise employs a coarse grained \emph{re-purposing of machines} and moves machines between the prompt and token pools.
Re-purposing of machines is done infrequently, typically only if they stay in the mixed pool for a considerable amount of time.

\myparagraph{Request routing}
CLS uses Join the Shortest Queue (JSQ) scheduling~\cite{jsq1,jsq2} to assign a prompt and a token machine to each request.
Queue lengths are defined by the number of pending tokens.
Each machine regularly communicates to the CLS changes in its memory capacity or pending queue.
Note that this does not happen at every iteration boundary.
We simultaneously assign both the prompt and token machine when scheduling requests, since we can then overlap KV-cache transfers with prompt computation to reduce transfer overheads (\Cref{sec:designkv}).

When routing requests, if the pending queue is bigger than a certain threshold, the CLS looks for target machines in the mixed pool.
If the mixed pool is also full, it proceeds to look in the opposite pool (\ie, a token machine to run prompts and vice versa) and moves the machine into the mixed pool.
Machines in the mixed pool operate exactly as a non-Splitwise machine would, with mixed batching.
Once the queue of mixed requests is drained, the CLS transitions the machine back to its original pool.
For example, when the queue is too long, we can move a prompt machine to the mixed pool to run tokens; once the machine is done running tokens, we transition the machine back into the prompt pool.

\subsection{Machine-level scheduling}
The MLS runs on each machine and is responsible for tracking the GPU memory utilization, maintaining the pending queue \circled{4}, deciding the batch for each iteration, and reporting the relevant status to the CLS.

\myparagraph{Prompt machines}
The MLS simply uses first-come-first-serve (FCFS) to schedule prompts.
The results in \Cref{fig:char_tput_prompt} show that after 2048 prompt tokens, the throughput degrades.
For this reason, the MLS restricts the batching of multiple prompts together to 2048 tokens in total.
This is a configurable value, and can change for a different model or hardware.

\myparagraph{Token machines}
The MLS uses FCFS to schedule tokens and batches as much as possible.
\Cref{fig:char_tput_token} shows that the token generation throughput keeps scaling up with the batch size until the machine runs out of memory.
For this reason, the MLS tracks the memory and starts queueing tokens once the machine is close to running out of memory.

\myparagraph{Mixed machines}
To meet the TTFT SLO, the MLS must prioritize running prompts and schedule any new prompts in the pending queue immediately.
If the machine is running token phases and has no additional capacity to run the prompt phase, the MLS will \emph{preempt} tokens.
To avoid \emph{starvation} of the token phase due to preemption, we increase the priority of the token with age and limit the number of preemptions that each request can have.

\subsection{KV-cache transfer}
\label{sec:designkv}
As discussed in \Cref{sec:background}, the KV-cache is generated during the prompt phase of the request, and it continuously grows during the token generation phase.
In Splitwise, we need to transfer the KV-cache from the prompt machine to the token machine \circled{5} (shown in \Cref{fig:systemdiag}) to complete the inference.
This transfer delay is the main overhead associated with Splitwise.
In this section, we discuss the impact of KV-cache transfer and how we optimize it.

\begin{figure}
    \centering
     \subfloat[Serialized KV-cache transfer.]{
        \includegraphics[width=0.46\columnwidth]{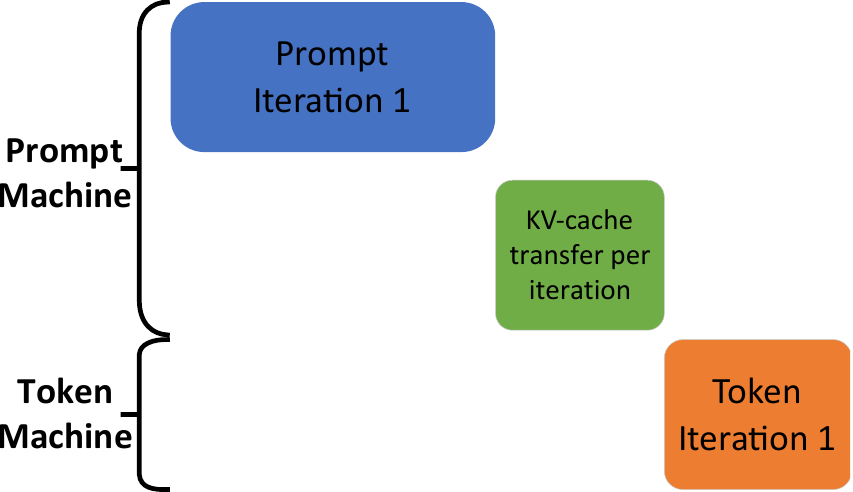}
        \label{fig:naivekv}
    }
    \subfloat[Optimized KV-cache transfer per-layer during prompt phase.]{
        \includegraphics[width=0.46\columnwidth]{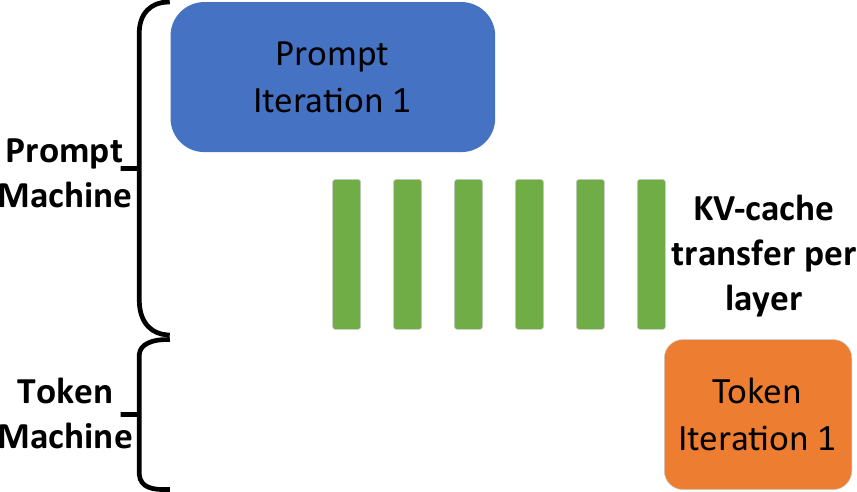}
        \label{fig:gantt_overlap}
    }
    \caption{Optimizing KV-cache transfer in Splitwise.}
    \label{fig:kvcachedesign}
\end{figure}

\Cref{fig:naivekv} shows the Gantt chart for the prompt phase, the KV-cache transfer, and the token generation phase for a single batch of requests when naively transferring the KV cache in a serialized way.
The KV-cache transfer starts only after the prompt phase has finished and the first token is generated.
Further, it needs to complete before the next output token can be generated in the token generation phase.
This directly impacts the maximum TBT and end-to-end latency of inference.

The time required for the transfer depends on the size of the KV cache (which is directly proportional to the number of prompt tokens) and on the bandwidth of the interconnect between the prompt and the token machines.
Even when using fast InfiniBand links, the \rev{transfer} overhead for large prompt sizes could become a significant fraction of the TBT.

In Splitwise, we optimize the KV-cache transfer by overlapping it with the computation in the prompt phase.
As each layer in the LLM gets calculated in the prompt machine, the KV cache corresponding to that layer is also generated.
At the end of each layer, we trigger an asynchronous transfer of the KV-cache for that layer while the prompt computation continues to the next layer.
\Cref{fig:gantt_overlap} shows this asynchronous transfer which reduces the \rev{transfer} overheads.
Layer-wise transfer also \rev{enables} other optimizations, such as earlier start of the token phase in the token machines, as well as earlier release of KV-cache memory on the prompt machines.

\rev{
Layer-wise KV-cache transfer happens in parallel with the prompt computation for the next layer.
This requires fine-grained synchronization per layer for correctness. 
Thus, it is possible to incur performance interference and increase the TTFT, especially for smaller prompts.
However, for small prompts the total KV-cache size is small and does not need the layer-wise transfer to hide the latency.
Since the number of tokens in a batch is already known at the start of computation, Splitwise picks the best technique for KV-cache transfer.
It uses serialized KV-cache transfer for smaller prompts and layer-wise transfer and for larger prompts.
We show that the overall transfer and interference overheads are relatively small in~\Cref{sec:expresults}.
}


\begin{table*}[]
\centering
\footnotesize
\begin{tabular}{@{}cccccccc@{}}
\toprule
                      &\multicolumn{3}{c}{\textbf{Prompt Machine}} & \multicolumn{3}{c}{\textbf{Token Machine}}  &\textbf{Prompt-Token}\\
                      & \textbf{Type}   &\textbf{Cost}  &\textbf{Power}     & \textbf{Type}   &\textbf{Cost}  &\textbf{Power} &\textbf{Interconnect Bandwidth} \\\midrule                  
\textbf{Splitwise-AA}         &DGX-A100     & 1$\times$     & 1$\times$ &DGX-A100     & 1$\times$     & 1$\times$     & 1$\times$                \\
\textbf{Splitwise-HH}          &DGX-H100     & 2.35$\times$     & 1.75$\times$ &DGX-H100     & 2.5$\times$     & 1.75$\times$  & 2$\times$    \\
\textbf{Splitwise-HHcap}       &DGX-H100     & 2.35$\times$     & 1.75$\times$ &DGX-H100     & 2.5$\times$     & 1.23$\times$  & 2$\times$   \\
\textbf{Splitwise-HA}      &DGX-H100     & 2.35$\times$     & 1.75$\times$ &DGX-A100     & 1$\times$     & 1$\times$  & 1$\times$   \\
\bottomrule
\end{tabular}
\caption{Evaluated Splitwise designs all normalized to DGX-A100}
\label{tab:systems_evaluated}
\end{table*}

\subsection{Provisioning with Splitwise}
\label{sec:provision}
We leverage Splitwise to optimize LLM inference cluster deployments for power, cost, and throughput.

\myparagraph{Type of machines}
We propose four main variants of Splitwise-based systems:
\emph{Splitwise-AA},
\emph{Splitwise-HH},
\emph{Splitwise-HA},
and \emph{Splitwise-HHcap}.
The nomenclature is simply drawn from the first letter representing the Prompt machine type, and the second letter representing the Token machine type.
``A'' represents a DGX-A100 machine, ``H'' represents a DGX-H100 machine,
and ``Hcap'' represents a power-capped DGX-H100 machine.
\Cref{tab:systems_evaluated} shows a summary of the cost, power, and hardware in each of our evaluated systems.

Splitwise-AA uses DGX-A100 for both prompt and token pools, while Splitwise-HH uses DGX-H100 for both.
These two variants represent the commonly available setups in providers where machines are homogeneous and interchangeable.

Splitwise-HA uses DGX-H100 for the prompt pool and DGX-A100 for the token pool.
We choose this configuration based on \Cref{tab:a100h100results}, and the Insight \rom{7} (\ie, A100s can be more cost- and power-efficient for the token phase).

Splitwise-HHcap uses DGX-H100 machines for both prompt and token pools.
However, we power cap the token machines down to 70\% of their rated power, with each GPU capped by 50\% of the power.
We propose this design based on \Cref{fig:char_powercap} and Insight \rom{7} (\ie, the prompts phase is impacted by power caps while token has no performance impact with 50\% lower power cap per GPU).

\myparagraph{Number of machines}
The LLM inference cluster deployment must be sized with the appropriate number of prompt and token machines.
Our methodology involves searching the design space using our event-driven cluster simulator, which is described in detail in \Cref{sec:method}.
We need to provide as input:
(1) the target cluster design (\eg, Splitwise-HA or Splitwise-HHcap),
(2) an LLM-specific performance model that can estimate the TTFT and TBT at various input, output, and batch sizes,
(3) a short trace derived from the target prompt and token size distributions for the service (\eg, \Cref{fig:prompt_token_sizes}), 
(4) the SLOs (\eg, \Cref{tab:eval-slo}),
(5) the constraints (\eg, throughput),
and (6) the optimization goal (\eg, minimize cost).
Using this information, our provisioning framework searches the space for the desired optimal point.
For example, searching with a throughput constraint and a cost minimization goal gives us iso-throughput cost-optimized clusters across different designs. 

\begin{figure}
    \centering
    \includegraphics[width=1.0\columnwidth]{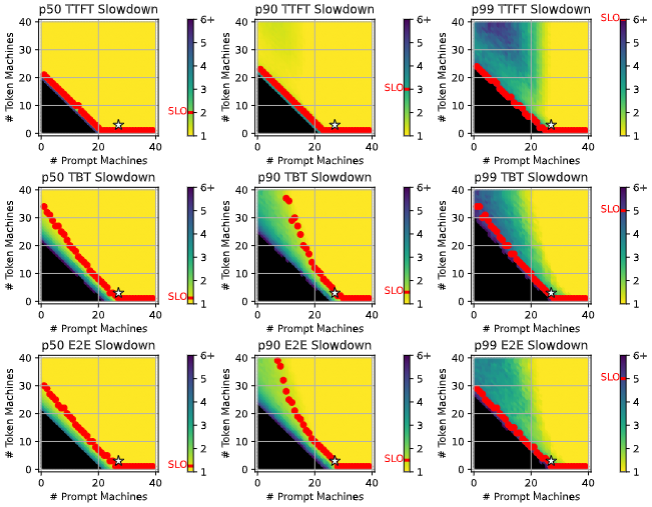}
    \caption{Design space for provisioning a Splitwise-HH cluster.
    Cluster configurations targets a peak throughput of 70 RPS.
    The cost-optimal Splitwise-HH configuration is marked with $\star$ (27 prompt and 3 token machines).
    }
    \label{fig:provisioning}
\end{figure}

\myparagraph{Search space}
\Cref{fig:provisioning} shows an example of the two-dimensional search space for the number of prompt and token machines under Splitwise-HH for the coding workload (using a 2-minute trace).
The simulator outputs the various percentiles for TTFT, TBT, and E2E latencies.
Then, we select the clusters that meet the SLOs for each of these metrics and optimize our target function.
For example, \Cref{fig:provisioning} shows a $\star$ for the setup with 27 prompt and 3 token machines with the lowest cost that achieves 70 RPS.
We call this setup \emph{iso-throughput cost-optimized}.

\myparagraph{Optimization}
We can use three optimization goals:
\emph{throughput}, \emph{cost}, and \emph{power}.
Throughput optimization is important for both, the cloud service provider (CSP) and the user.
Cost optimization has different importance levels to the CSP and the user.
For the CSP, a higher cost for the same throughput might be acceptable if there are gains in power and space requirements for the cluster.
However, for the end-user, a higher cost at the same throughput is generally unacceptable.
Finally, power optimization is attractive for a CSP, since it enables more GPUs to be deployed in the same datacenter~\cite{patel2023polca,patel2024llmpower}, but it may not be as important to the user.
We only consider the provisioned power, and not the dynamic power utilization, in our study.

\rev{
\subsection{Practical Considerations}
\label{sec:splitwise_considerations}

\myparagraph{Accuracy impact}
Splitwise does not impact accuracy since it uses lossless KV-cache transfer and does not add any randomization.
It executes inference with the same parameters and state as on a single machine.

\myparagraph{Scalability}
Since LLM requests are much longer than typical ML requests~\cite{gujarati2020_Serving,gupta2020deeprecsys}, they incur lower scheduling overhead for similar cluster sizes. 
However, the CLS may become a scalability bottleneck for large clusters.
Insights from prior work on partitioned or replicated scheduling could help improve scalability~\cite{ousterhout2013_Sparrow,boutin2014apollo, schwarzkopf2013omega} and are orthogonal to Splitwise.

\myparagraph{Reliability and fault tolerance}
If the prompt or the token machine fail, Splitwise simply restarts requests from scratch, similar to today’s LLM serving systems~\cite{vllm,huggingface_tgi}.
Alternatively, Splitwise could checkpoint the KV-cache generated after prompt computation into an in-memory database.
To recover, Splitwise can use this cache to skip prompt recomputation, and start right away with the token phase.
The KV-cache could also be checkpointed periodically during the token phase.
Designing safe and efficient failure recovery is out of scope for our paper.
}

\section{Methodology}
\label{sec:method}

\subsection{Experimental setup}
\label{sec:expt_setup}

To evaluate our proposal on real hardware, we implement Splitwise's KV-cache transfer mechanism on top of vLLM~\cite{vllm}. Our implementation is open source~\cite{splitwise_vllm_pr}.
We run this modified vLLM on two DGX-A100 and two DGX-H10 virtual machines (VMs) on Microsoft Azure with specifications from \Cref{tab:a100_vs_h100_specs}.
These are the VMs used to collect the characterization data in \Cref{sec:motivation}.
These machines are connected with InfiniBand and the DGX-H100s have double the bandwidth (\ie, 400 Gbps).

Since vanilla vLLM only supports continuous batching with token preemption which can lead to much higher TBT, we implement state-of-the-art mixed continuous batching~\cite{yu2022orca} as discussed earlier in \Cref{fig:batching-gantt-chart}(c).

Our implementation of the Splitwise technique assigns machines either a prompt role, or a token role.
As the prompt machine generates the first token, it transfers the KV-cache to the token machine using the technique described in \Cref{sec:designkv}.
We use MSCCL++~\cite{mscclpp}, an optimized GPU-driven communication library, to implement the naive and layer-wise KV cache transfers.

\rev{
In our implementation, the prompt machine uses the zero-copy one-sided \texttt{put} primitive of MSCCL++ to send KV-cache data over InfiniBand as soon as it is ready, without requiring the token machine to issue any receive instructions.
Once we have issued a \texttt{put} for all layers, the prompt machine signals a semaphore that the token machine waits on.
The synchronization done with the help of semaphores uses the same InfiniBand connection used to send KV-cache data.
When processing a batch of prompts, each request is assigned a different semaphore since it may be routed to different token machines.
We ship the KV-caches block-by-block in vLLM.
To minimize the number of transfers, we also consider the contiguity of KV blocks as long as they use the same semaphore.
}


\subsection{Simulator setup}
\label{sec:simulator}

We build a simulator to explore cluster designs and evaluate Splitwise at scale.
The simulator code is open source~\cite{splitwise_sim}.


\begin{figure}[t]
    \centering
    \includegraphics[width=0.75\columnwidth]{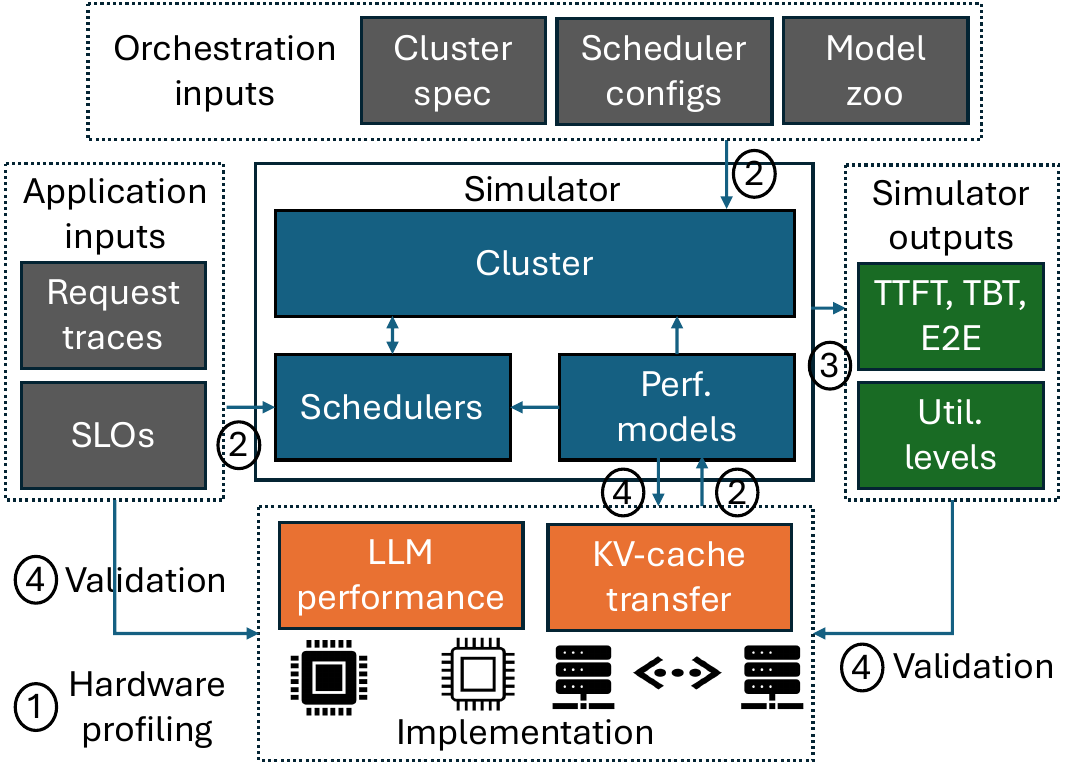}
    \caption{\rev{Overview of the design of the Splitwise simulator.}}
    \label{fig:simulator}
\end{figure}

\rev{
\Cref{fig:simulator} shows the design of our simulator.
The simulator is event-driven and faithfully models the Splitwise machine pools, schedulers, machine-level memory and queues, and KV-cache transfer.
We first profile the LLM on the target hardware with various input/output sizes~\circled {1}.
Based on the characterization profiles, we build a performance model.
The simulator takes as input the request traces, SLOs, the performance model, and the configurations for cluster and scheduler~\circled {2}.
For our evaluation, we use the prompt and token size distributions from the production traces in \Cref{sec:motivation}.
We tune the Poisson arrival rate to increase and decrease the load (requests per second) for cluster sizing.
The simulator provides the achieved metrics per request (TTFT, TBT, E2E), and the machine utilization levels~\circled {3}.
We cross-validated the performance model with hardware experiments to ensure accuracy; we also validated the simulator end-to-end using production load with over 50K iterations to ensure fidelity~\circled {4}.
}

\rev{
\myparagraph{Performance model}
We build a piece-wise linear performance model using performance profiles at various batch sizes, input sizes, output sizes, in the required parallelism configuration on A100 and H100 machines from~\Cref{sec:motivation}.
We validate that our performance model has high accuracy; it incurs a mean absolute percentage error (MAPE) of less than 3\% when evaluated with a 80:20 train:test dataset split.

\myparagraph{Communication model}
In our evaluation, KV-cache transfers cause inter-machine communication, whereas tensor parallelism only causes intra-machine communication.
We model inter-machine communication overheads by benchmarking our KV-cache transfer implementation over Infiniband in~\Cref{sec:expresults}.
}

\myparagraph{SLOs}
To determine the maximum throughput that can be supported by a given cluster design, we use P50, P90, and P99 SLOs for TTFT, TBT, and E2E latency metrics.
\Cref{tab:eval-slo} shows our SLO definition using DGX-A100 as a reference.
We require all nine SLOs to be met.
SLOs on TTFT are slightly looser, since it has a much smaller impact on the E2E latency.

\begin{table}[t]
\centering
\begin{tabular}{cccc}

\toprule
            & \textbf{P50} & \textbf{P90} & \textbf{P99} \\
     \midrule
     \textbf{TTFT}   & 2$\times$    & 3$\times$   & 6$\times$\\
     \textbf{TBT}    & 1.25$\times$ & 1.5$\times$ & 5$\times$\\
     \textbf{E2E}    & 1.25$\times$ & 1.5$\times$ & 5$\times$\\
\bottomrule
\end{tabular}
\caption{SLO expressed as slowdown compared to a request running on DGX-A100 under no contention.}
\label{tab:eval-slo}
\end{table}

\myparagraph{Baselines}
We compare our Splitwise designs against Baseline-A100 and Baseline-H100.
The clusters in these baselines consist of just DGX-A100s and DGX-H100s, respectively.
Both baselines use the same mixed continuous batching that Splitwise uses for mixed pool machines (described in \Cref{sec:scheduler}).

\section{Evaluation}
\label{sec:eval}

\subsection{Experimental results}
\label{sec:expresults}

\begin{figure}[t]
    \centering
    \includegraphics[width=0.6\columnwidth]{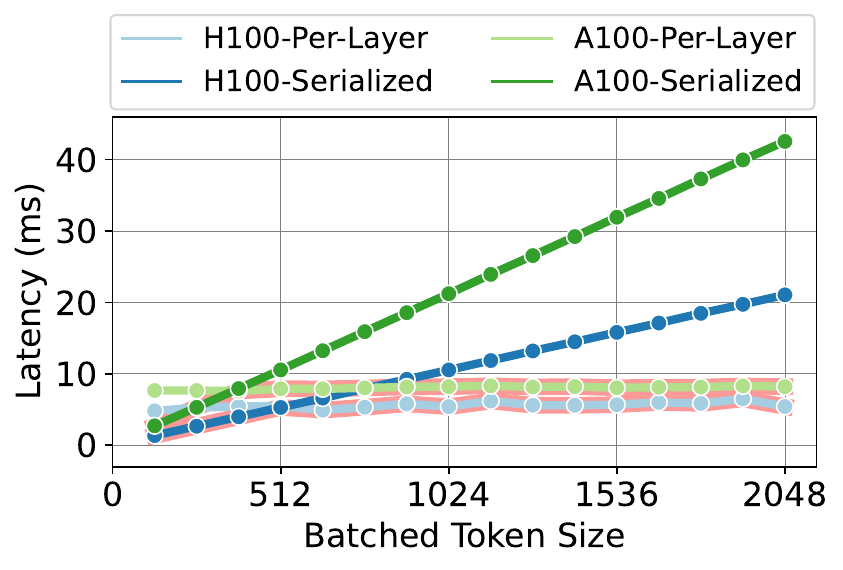}
    \caption{Overhead of the KV-cache transfer as the prompt size increases on A100s and H100s.}
    \label{fig:kvcacheperf}
\end{figure}

\myparagraph{KV-cache transfer latency}
We first measure the latency to transfer the KV-cache as the prompt size grows.
\Cref{fig:kvcacheperf} shows the visible transfer latency on both A100 and H100 setups with the naive and optimized transfer design as discussed in \Cref{fig:kvcachedesign}.
Compared to the prompt computation time, the overhead is minimal ($<7\%$).
The time for serialized transfers linearly increases with the prompt size since the size of the KV-cache also increases.
The optimized per-layer transfer, on the other hand, hides much of the latency.
For these transfers, we see a constant non-overlapped transfer time of around 8ms for the A100 and around 5ms for the H100 setup.
The H100 setup has double the bandwidth of the A100 setup (\ie, 200 vs 400 Gbps), and the impact of this can be clearly seen with transfers in the H100 setup happening about twice as fast as those in the A100 setup.

As discussed in \Cref{sec:designkv}, for small prompt sizes ($<512$ in H100), Splitwise uses the serialized KV-cache transfer and for larger prompts, it uses per-layer transfers.


\begin{figure}[t]
    \centering
    \includegraphics[width=0.7\columnwidth]{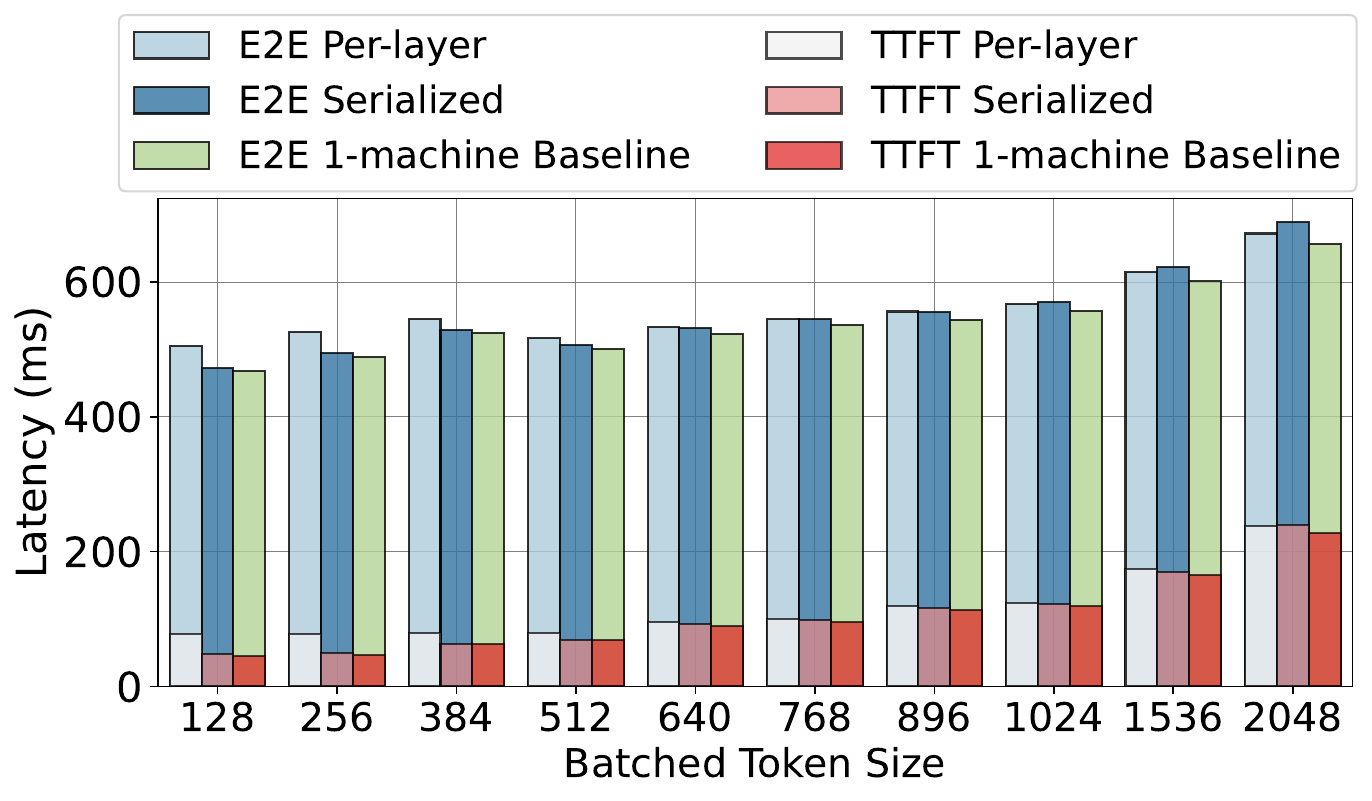}
    \caption{Overhead of KV cache transfer on TTFT, E2E latency for coding trace for A100 and H100.}
    \label{fig:kvcacheperfe2e}
\end{figure}

\myparagraph{End-to-end impact}
Next, we run the coding trace on the 2-machine Splitwise setups without batching, and compare the observed latency metrics to a 1-machine baseline setup with no batching.
\Cref{fig:kvcacheperfe2e} shows our results.
The latency impact of serially transferring the KV-cache grows up to 3\% of the E2E with large prompts.
However, Splitwise only incurs 0.8\% of E2E.
In a user-facing inference, the only visible impact of KV-cache transfer overhead is the latency for the second token.
Splitwise adds a 16.5\% latency to the second token, as compared to the 64\% overhead from a serialized transfer.
Overall, the transfer impact in Splitwise is hardly perceivable even in a user-facing inference. 

\subsection{Iso-power throughput-optimized clusters}

\myparagraph{Cluster provisioning}
We provision clusters using the methodology described in \Cref{sec:provision}.
We target a specific workload (\eg, conversation) at a peak load with the same power (\ie, iso-power) for each cluster design.
For the baseline, we use the power for 40 DGX-H100 machines as our target peak power.
For the A100 baseline, we can fit 70 DGX-A100 machines under the same power budget.
We denote these two designs as 40P/T and 70P/T respectively, since they both use mixed batching in all machines.

\begin{figure}[t]
    \centering
    \subfloat[Coding trace.]{
    \includegraphics[width=1.0\columnwidth]{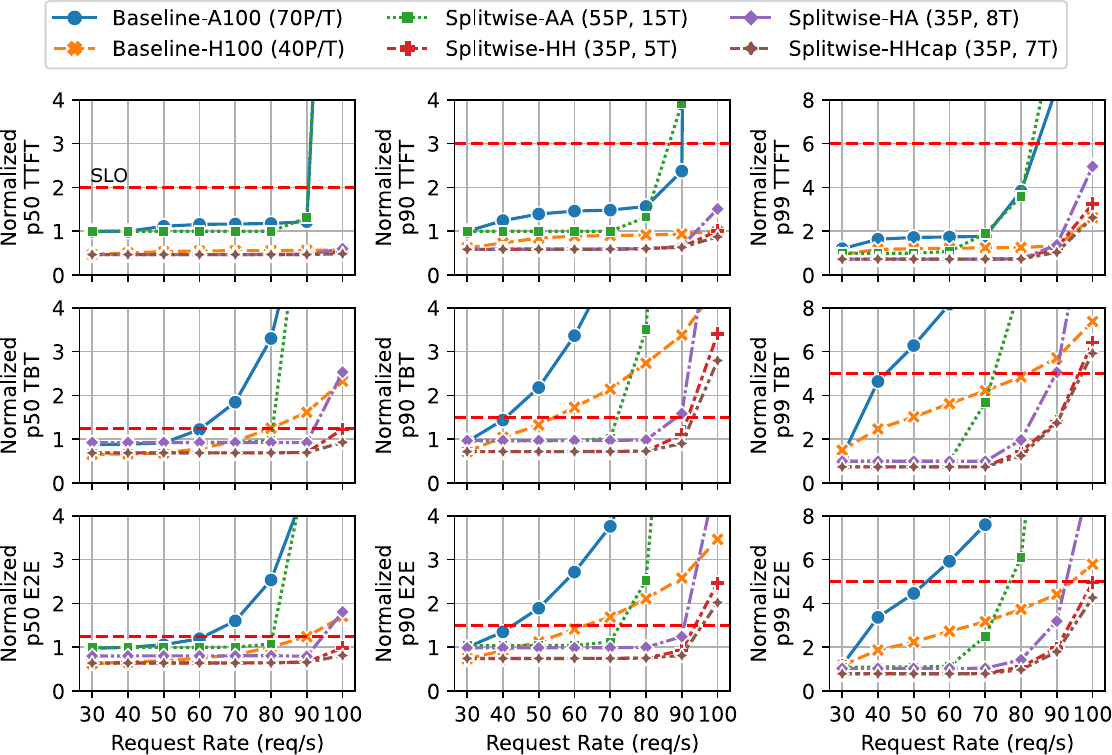}
    \label{fig:isopowerfullcode}
    }
    \vspace{2pt}
    \subfloat[Conversation trace.]{
    \includegraphics[width=1.0\columnwidth]{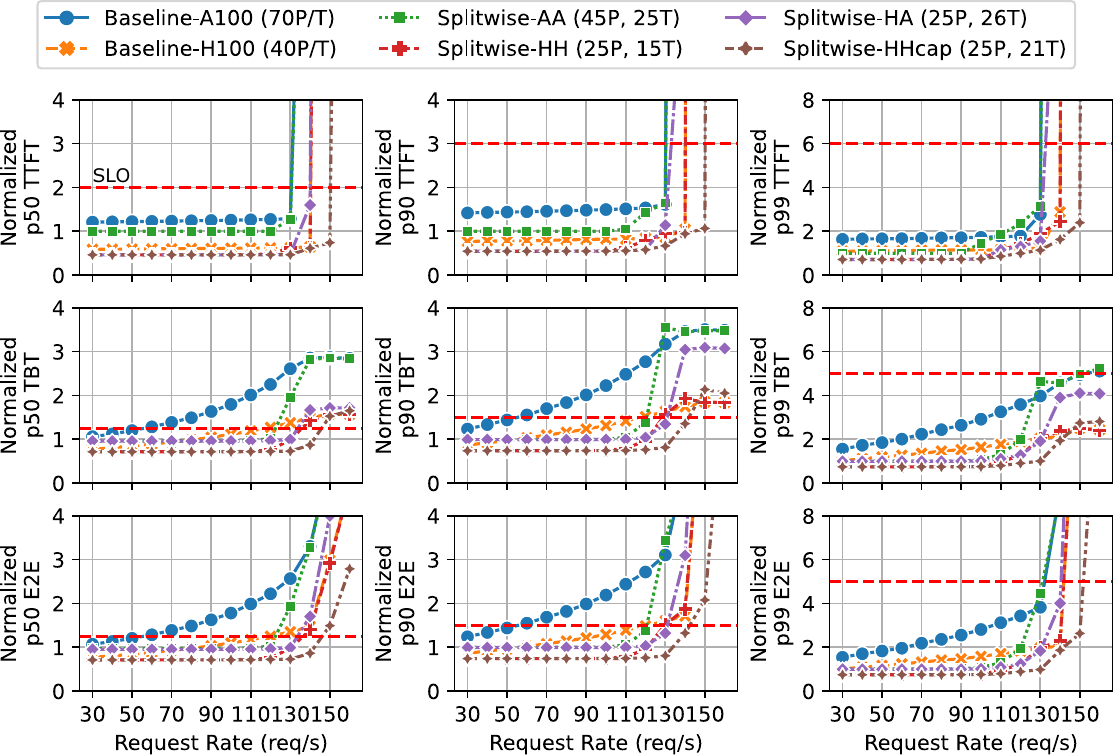}
    \label{fig:isopowerfullconv}
    }
    \caption{Latency metrics across input loads for iso-power throughput optimized clusters.
    Dashed red lines indicate SLO.
    }
    \label{fig:isopowerfull}
\end{figure}

For Splitwise cluster designs under the coding trace, Splitwise-AA provisions 55 prompt machines and 15 for the token pool, denoted as (55P, 15P).
Note that like Baseline-A100, Splitwise-AA also provisions 75\% more machines than Baseline-H100.
The legends in \Cref{fig:isopowerfull} show the different provisioning choices under coding and conversation workloads.
Request size distributions reflect in the machine pool sizing.
For example, we provision more prompt machines under Splitwise-HH (35P, 5T) for the coding trace, while we provision more token machines (25P, 15T) for the conversation trace.

\myparagraph{Latency and throughput}
\Cref{fig:isopowerfull} shows a deep dive into all the latency metrics at different input load for each cluster design with the same power (\ie, iso-power).
For the coding trace (\Cref{fig:isopowerfullcode}), Splitwise-HH, Splitwise-HHcap, and Splitwise-AA all perform better than Baseline-H100.
As the load increases, Baseline-H100 suffers from high TBT due to mixed batching with large prompt sizes.
Although Splitwise-AA can support higher throughput, its TTFT is consistently higher than most designs.
Splitwise-HA clearly bridges the gap by providing low TTFT and E2E at high throughput.
The mixed machine pool in Splitwise becomes useful at higher loads to use all the available hardware without fragmentation.
This benefit can be seen clearly in the P50 TBT chart for Splitwise-HA, where after 90 RPS, H100 machines jump into the mixed machine pool and help reduce TBT. 

For the conversation trace (\Cref{fig:isopowerfullconv}), Splitwise-HHcap clearly does better on all fronts, including latency.
This is because its token generation phases typically run for much longer than in the coding trace, which is beneficial for the token machines.

\begin{figure}[t]
    \centering
     \subfloat[Low load (70 RPS).]{
        \includegraphics[width=0.48\columnwidth]{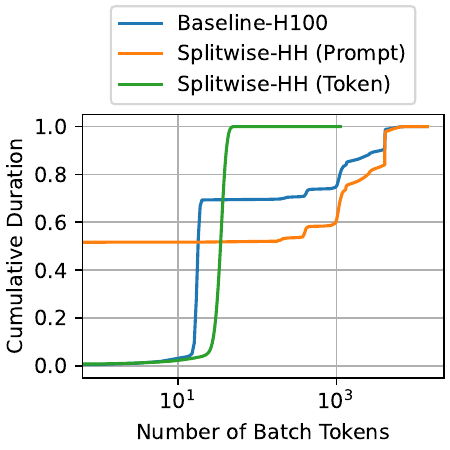}
        \label{fig:batchlowtput}
    }
    \subfloat[High load (130 RPS).]{
        \includegraphics[width=0.48\columnwidth]{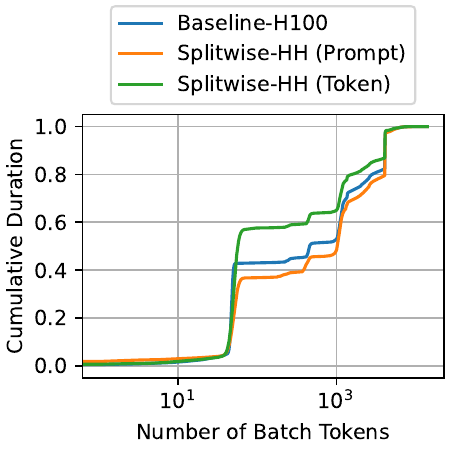}
        \label{fig:batchhightput}
    }
    \caption{Cumulative distribution of time spent at various batched token sizes for iso-power throughput-optimized design.
    }
    \label{fig:batchingeval}
\end{figure}

\myparagraph{Impact on batched tokens}
\Cref{fig:batchingeval} shows the cumulative distribution of time spent processing a varying number of batched active tokens in an iso-power throughput-optimized cluster. The distributions are collected by running the conversation trace at low (70 RPS) and high (130 RPS) loads.

At low load, all 40 Baseline-H100 machines spend 70\% of the time running $\leq$15 tokens, and the rest running mixed batches with large prompts, which affects TBT and E2E.
The 35 Splitwise-HH prompt machines are mostly idle, and when active, run much larger batches of tokens.
The 15 Splitwise-HH token machines also do a better job at batching.
Overall, Splitwise machines have better batching and latency at 70 RPS.
At high load, since the mixed pool is utilized more, the batch sizes start looking similar across prompt and token machines.

\begin{figure}[t]
    \centering
     \subfloat[Iso-power.]{
        \includegraphics[width=0.48\columnwidth]{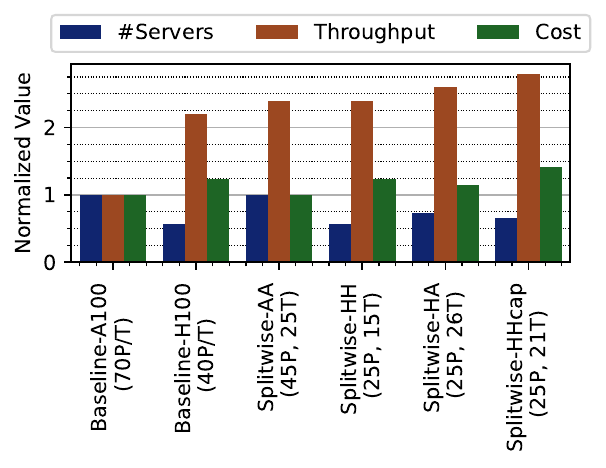}
        \label{fig:isopowertput}
    }
    \subfloat[Iso-cost.]{
        \includegraphics[width=0.48\columnwidth]{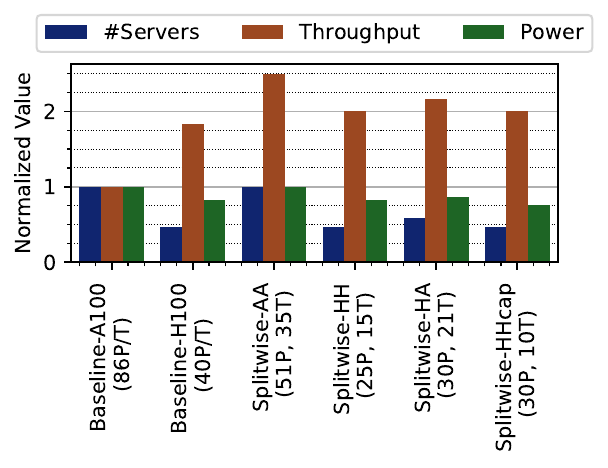}
        \label{fig:isocosttput}
    }
    \caption{Summary of throughput-optimized cluster designs.}
    \label{fig:throughputopt}
    \vspace*{-0.5cm}
\end{figure}

\myparagraph{Summary plot}
\Cref{fig:isopowertput} summarizes the results across all cluster metrics for iso-power throughput-optimized designs for the conversation trace.
We use Baseline-A100 as the baseline.
Compared to Baseline-A100, Splitwise-AA delivers 2.15$\times$ more throughput at the same power and cost.
Splitwise-HA delivers 1.18$\times$ more throughput at 10\% lower cost and the same power.

\begin{figure}[t]
    \centering
     \subfloat[Power-optimized.]{
        \includegraphics[width=0.48\columnwidth]{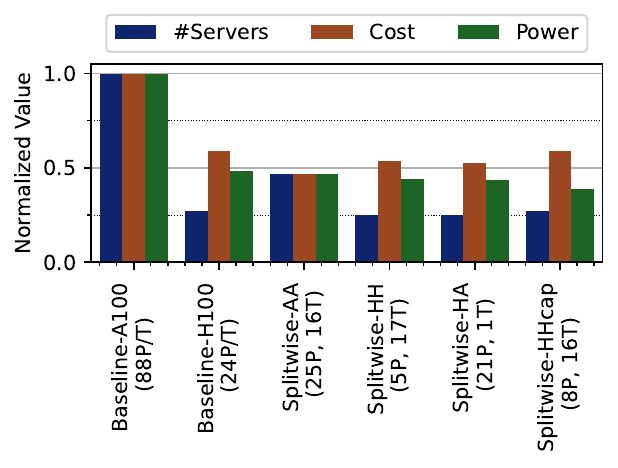}
        \label{fig:isotputpower}
    }
    \subfloat[Cost-optimized.]{
        \includegraphics[width=0.48\columnwidth]{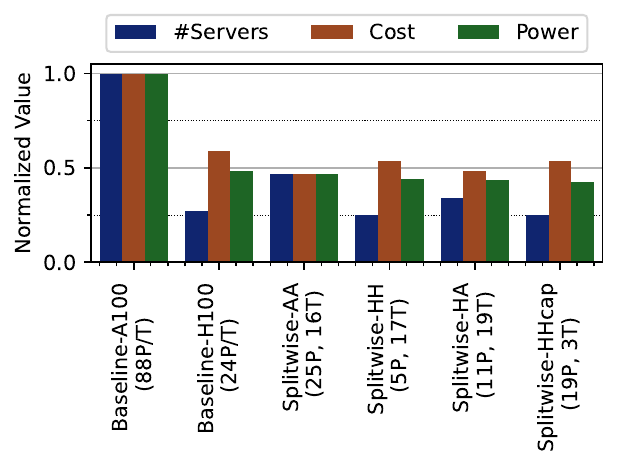}
        \label{fig:isotputcost}
    }
    \caption{Summary of iso-throughput cluster designs.}
    \label{fig:isotput}
\end{figure}

\subsection{Other cluster optimizations}
We have described \emph{iso-power throughput-optimized clusters} in detail.
For the rest of the cluster optimization evaluation, we only discuss the summary plots.

\myparagraph{Iso-cost throughput-optimized}
\Cref{fig:isocosttput} shows the summary plot for iso-cost clusters, with their space, throughput, and power requirements.
We find that Splitwise-AA provides the best throughput for the same cost, namely 1.4$\times$ more throughput than Baseline-H100, running at 25\% more power, and 2$\times$ the space.
This is an interesting operational point for most customers who may not care about power and space, instead preferring the 40\% higher throughput using older, more easily available GPUs.
In contrast, the preferable choice for the CSP is less clear.

\myparagraph{Iso-throughput power-optimized}
\Cref{fig:isotputpower} shows cluster designs that yield same throughput at the least power.
Splitwise-HHcap can achieve the same throughput as Baseline-H100 at 25\% lower power at the same cost and space.
This can be a clear win for the CSPs.

\myparagraph{Iso-throughput cost-optimized}
\Cref{fig:isotputcost} shows the cost-optimized versions of the iso-throughput design.
Note that there are no changes to any of the homogeneous designs between \Cref{fig:isotputpower,fig:isotputcost}.
This is because the prompt and token machines have the same cost and power.
However, Splitwise-HA and Splitwise-HHcap arrive at slightly different results with the cost and power optimizations.
\Cref{fig:isotputcost} shows that with Splitwise-AA, customers can achieve the same throughput as Baseline-H100 at 25\% lower cost.

\subsection{Impact of workload changes}
\label{subsec:exp-workload-changes}
\begin{figure}
    \centering
     \subfloat[Conversation trace running on a cluster designed for coding.]{
        \includegraphics[width=1.0\columnwidth]{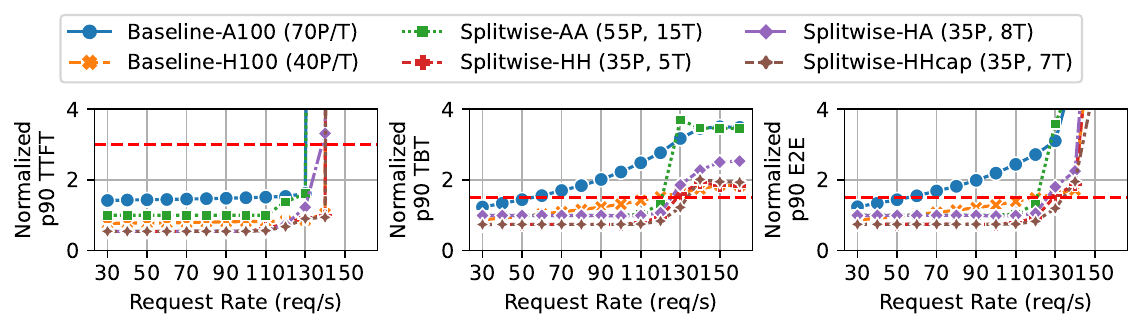}
        \label{fig:convoncode}
    }
    \vspace{0pt}
    \subfloat[Llama-70B, on a cluster designed for BLOOM-176B.]{
        \includegraphics[width=1.0\columnwidth]{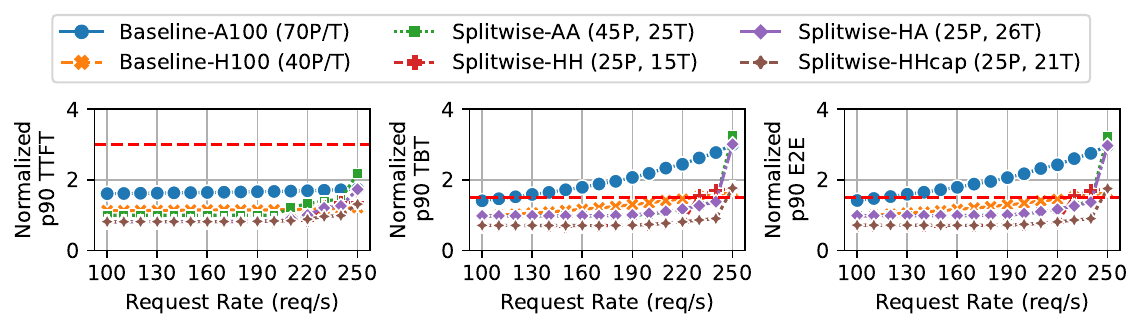}
        \label{fig:llamaonbloom}
    }
    \caption{Latency impact of running a workload on a cluster designed for another workload. Dashed red lines indicate SLO.}
    \label{fig:switcharoo}
\end{figure}

So far, we have tested a trace and a model on clusters optimized for a specific workload pattern and model.
To test the Splitwise' robustness, we now run conversation trace on a cluster meant for coding service, and Llama-70B on a cluster meant for BLOOM-176B.
\Cref{fig:switcharoo} shows these results for iso-power throughput-optimized clusters.

\myparagraph{Changing workload trace}
Compared to \Cref{fig:isopowerfullconv}, we find that in \Cref{fig:convoncode}, the Baseline clusters are similarly sized and incur no throughput or latency impact.
Splitwise-AA and Splitwise-HH with the mixed pool morph well to meet the requirements of the new workload, and they see no throughput or latency impact.
Since Splitwise-HA and Splitwise-HHcap have different types of machines in the prompt and token pools, they experience a throughput setback of 7\% from the respective cluster optimized designs for conversation trace. Note that all the Splitwise designs still perform much better than any of the Baseline designs. 

\myparagraph{Changing model}
\Cref{fig:llamaonbloom} shows that Llama-70B can support much higher throughput in the same cluster design than BLOOM-176B, given its fewer parameters (\Cref{tab:models}).
All the Splitwise designs out-perform both the Baseline designs at higher load.
Furthermore, Splitwise-HH and Splitwise-HHcap consistently achieve the best latency, even as the load increases.

\myparagraph{Summary}
Based on these two experiments, we conclude that Splitwise can morph according to the requirements of the workload using its smart scheduling, and it is robust to changes in the LLMs, request load, and token distributions.

\subsection{Cluster design for batch job}
We design various clusters with Splitwise under strict latency SLOs, even when we are optimizing for throughput.
This is unnecessary for batch jobs, which can be stressed to high load for a high token generation throughput.
We find that upon stressing our iso-power throughput-optimized clusters, Baseline-A100 and Splitwise-AA have the best throughput per cost at 0.89 RPS/\$.
At high load, Splitwise devolves into the iso-count Baseline, since it starts mixed batching with all the machines in the mixed pool.
The same holds true for Splitwise-HH and Baseline-H100, which achieve 0.75 RPS/\$.

\section{Discussion}
\label{sec:discussion}

\rev{
\myparagraph{Extensibility to new models}
Despite the plethora of model sizes from 2B parameters ~\cite{phi2,zhang2022opt} to 176B parameters~\cite{scao2022bloom} or more~\cite{chatgptapi}, all modern transformer-based generative LLMs have the distinct prompt processing and token generation phases. 
Similarly, even modifications and flavors like Mixture-of-Experts (MoEs) have these phases.
Since Splitwise is built solely by exploiting these phases, it is applicable to all of the current and upcoming LLMs, as long as the auto-regressive nature of the workload requires these two phases.
Note that as shown in \Cref{subsec:exp-workload-changes}, clusters provisioned with Splitwise for one model can also efficiently serve other models. 
}

\rev{
\myparagraph{Alternative compute hardware}
In this work, we use NVIDIA H100 and A100 GPUs since they are commonly used for LLM inference in datacenters today~\cite{nvidia_hopperincloud}.
Smaller datacenter GPUs like NVIDIA T4 lack enough memory to run modern LLMs efficiently.
In general, our methodology is applicable to any hardware (including CPUs, FPGAs, ASICs~\cite{davidpatterson2019_Domain}) that aligns with the computational requirements of prompt and token phases.
Our characterization suggests that prompt phases need high compute capability and memory bandwidth with low memory capacity, whereas token phases need moderate compute capability with high memory capacity and bandwidth.
Thus, GPUs like AMD MI-250~\cite{mi250} and CPUs like Intel Sapphire-Rapids (with HBM)~\cite{sprhbm} could be effective token machines.
Since we do not have access to such hardware and/or optimized LLM implementations, we leave this to future work.
}


\myparagraph{Interconnect between prompt and token machines}
In this work, we assume Infiniband connection between the prompt and token machines in all the designs (albeit, lower bandwidth when A100s were involved).
Although this is common for all homogenous machines, Splitwise-HA is not be readily available with an Infiniband connection between H100s and A100s, even though technically feasible.
The alternative could be HPC clouds, with Infiniband connections through the CPU~\cite{azurehpcvm}, or Ethernet, using RoCE~\cite{roce}.
Given our optimized KV-cache transfer that helps reduce critical latency, an interconnect with $10\times$ lower bandwidth would likely still be beneficial.
To further reduce our bandwidth utilization, we could also compress the KV-cache before transferring it across the network~\cite{liu2023deja}.

\myparagraph{Heterogeneous prompt/token machines}
Although Splitwise is robust to varied models and input traces, we recognize that fragmenting a data center with different types of GPUs (\eg, Splitwise-HA) may bring its own challenges for the CSP.

\myparagraph{Conversation back and forth}
Chat APIs for LLMs today require the user to send the complete context of the conversation so far~\cite{chatgptapi}.
However, in the future, services may have enough GPU capacity to cache the context and avoid recomputation.
This could sway the memory utilization pattern of the prompt phase from our characterization.
Furthermore, it may require transferring the KV-cache back to a prompt machine to be ready for the next conversation request.

\section{Related Work}
\label{sec:related}

\rev{

\myparagraph{Heterogeneous scheduling and dataflow systems}
Prior work has studied heterogeneous scheduling for a variety of interactive services~\cite{zhang2019_MArk,patel2023hybrid,romero2019_INFaaS}.
These works exploit hardware heterogeneity to strike a balance between different objectives such as cost, energy, and performance.
However, they run the entire workload on the same machine.
Research on heterogeneous multiprocessor CPU scheduling attempts to match workload heterogeneity to hardware heterogeneity~\cite{haque2015few,van2012scheduling,haque2017fasttoslow,kumar2003single,yang2017powerchief,chen2009efficient}.
These works use profiling or online monitoring with metrics like request length or hardware performance counters to identify workload phases and allocate them appropriately on heterogeneous processors.
However, they do not consider the complexities with batching.
%
Distributed dataflow systems orchestrate large-scale computational graphs and aim to provide general-purpose programmability~\cite{dean2008mapreduce,isard2007dryad,shvachko2010hadoop,zaharia2012resilient}.
LLM inference under Splitwise can be viewed as a static computational graph with two stages, so it could be implemented using distributed frameworks that provide efficient GPU abstractions~\cite{moritz2018ray}.
Splitwise differs from these works since it uses a specialized two-phase design for generative LLM inference and leverages phase-aware resource management with efficient batching. 

\myparagraph{Model serving systems}
LLM inference serving is a rapidly developing field, with several recent works optimizing batching~\cite{yu2022orca,vllm,alpaserve,sarathi,aminabadi2022deepspeed}, scheduling~\cite{vllm,pope2023efficiently,sheng2023fairness,wu2023fast,turbomind,hong2023flashdecoding++}, and memory usage~\cite{flashattention,vllm,dettmers2022llmint8,sheng2023flexgen}.
Prior work has also proposed using CPUs and lower compute capability devices for LLM serving~\cite{bigdlllm,numenta}.
These approaches use the same machine for both prompt and token phase.
With Splitwise, they could improve throughput and latency by splitting phases.

Prior work on video and ML serving focuses on scheduling model chains with data dependencies under latency constraints~\cite{hu2021scrooge,crankshaw2020inferline,romero2019_INFaaS,kannan2019_GrandSLAm,albahar2022schedtune}.
Such schedulers rely on model profiling to make efficient allocation decisions and manage requests across machines.
Recommendation system inference exhibits compute/memory heterogeneity both within and across models.
Prior work exploits such heterogeneity to selectively schedule requests between CPUs and accelerators~\cite{gupta2020deeprecsys,kwon2019tensordimm}, colocate models with complementary memory usage~\cite{choi2023hera}, and partition compute/memory on heterogeneous hardware resources~\cite{hwang2020centaur,jiang2021fleetrec}.
Similarly, Splitwise exploits the heterogeneity within LLM inference requests.
However, it uses different optimizations due to the differences in LLM workload characteristics and requirements.
}

\section{Conclusion}
\label{sec:conclusion}

We extensively characterized the prompt computation and token generation phases of LLM inference to draw out differences in their system utilization patterns.
Based on our insights, we designed Splitwise to separate these phases onto different machines and enable phase-specific resource management.
Using Splitwise, we explored cluster designs optimized for throughput, cost, and power, and showed that they perform well even as workloads change.
Splitwise clusters under performance SLOs achieve $1.76\times$ better throughput with 15\% lower power at the same cost, or $2.35\times$ better throughput with same the cost and power than existing designs.

\section*{Acknowledgements}
We thank the reviewers for their helpful feedback.
We thank Chetan Bansal, Srikant Bhardwaj, Suriya Kalivardhan, Ankur Mallick, Deepak Narayanan, and Amar Phanishayee for insightful discussions.
Pratyush Patel was partially supported by NSF CNS-2104548 and a research grant from VMware.


\bibliographystyle{IEEEtranS}
\bibliography{refs}
\appendix
\section{Artifact Appendix}
\label{sec:artifact}

\subsection{Abstract}

We open source critical components needed to evaluate Splitwise; these could be repurposed to also evaluate future LLM inference serving systems. Our artifact includes:
\begin{itemize}
    \item Production traces from two LLM inference services at Microsoft Azure.
    \item A prototype implementation of Splitwise's KV-cache transfer mechanism in vLLM~\cite{vllm}.
    \item SplitwiseSim, a discrete event simulator to evaluate model serving in LLM inference clusters. 
\end{itemize}

Artifact functionality was only tested for the traces and SplitwiseSim due to limited hardware availability.

\subsection{Artifact check-list (meta-information)}

{\small
\begin{itemize}
  \item {\bf Data set: } Production traces available as a part of the artifact.
  \item {\bf Run-time environment: } Linux / Ubuntu.
  \item {\bf Hardware: } Two machines connected over GPU Infiniband for the vLLM prototype (\eg NVIDIA DGX-A100, NVIDIA DGX-H100). x86-64 CPU machine for SplitwiseSim.
  \item {\bf Publicly available?: } Yes.
  \item {\bf Code licenses (if publicly available)?: } MIT.
  \item {\bf Data licenses (if publicly available)?:} CC-BY.
  \item {\bf Archived (provide DOI)?:} 10.5281/zenodo.11003049.
\end{itemize}
}

\subsection{Description}

\myparagraph{How to access}
The entire artifact is available as an archive on Zenodo: \url{https://doi.org/10.5281/zenodo.11003049}.
Individual components are also available online as follows:
\begin{itemize}
    \item The production traces can be downloaded from the Azure Public Dataset GitHub repository~\cite{azure_llm_trace}. 
    \item The KV-cache transfer prototype can be downloaded from the vLLM GitHub repository, currently available as a pull request~\cite{splitwise_vllm_pr}.
    \item SplitwiseSim, and the associated experiment and plotting scripts, can be downloaded from a separate GitHub repository~\cite{splitwise_sim}.
\end{itemize}

\myparagraph{Hardware dependencies}
The KV-cache transfer prototype requires two GPU machines connected over Infiniband, such as NVIDIA DGX-A100s or NVIDIA DGX-H100s. SplitwiseSim requires a standard x86-64 CPU machine; multiple machines may be used to parallelize simulation runs.

\myparagraph{Software dependencies}
The KV-cache transfer prototype is built on top of vLLM~\cite{vllm} and MSCCL++~\cite{mscclpp}.
SplitwiseSim depends on a small set of publicly available Python packages, which can be installed via the included \texttt{requirements.txt}.

\myparagraph{Data sets} 
Coding and conversation traces from Microsoft Azure are available online as a part of the artifact release~\cite{azure_llm_trace}.



\subsection{Installation and Experiment Workflow}

Please refer to the README files within the artifact for installation and usage instructions.

%






\subsection{Methodology}

Submission, reviewing and badging methodology:

\begin{itemize}
  \item \url{https://www.acm.org/publications/policies/artifact-review-and-badging-current}
  \item \url{http://cTuning.org/ae/submission-20201122.html}
  \item \url{http://cTuning.org/ae/reviewing-20201122.html}
\end{itemize}

\end{document}